\documentclass[%
preprint,
 amsmath,amssymb,
 aps,
 prl,
 lengthcheck,%
]{revtex4-1}

\usepackage{graphicx}
\usepackage{url}
\usepackage[colorlinks=true,urlcolor=blue]{hyperref}

\def\eqns{\begin{eqnarray}}
\def\eqne{\end{eqnarray}}

\def\eqnsn{\begin{eqnarray*}}
\def\eqnen{\end{eqnarray*}}

\def\<{\langle}
\def\>{\rangle}

\def\eqref#1{~(\ref{#1})}

\def\pip2{PIP$_2$}

\begin{document}

\preprint{APS/123-QED}

\title{Electrodiffusion of Lipids on Membrane Surfaces}

\author{Y. C. Zhou}
\email{yzhou@math.colostate.edu}
\affiliation{Department of Mathematics, Colorado State University, Fort Collins, CO 80523, USA}%

\date{\today}

\begin{abstract}
Lateral translocation of lipids and proteins is a universal process on membrane surfaces.
Local aggregation or organization of lipids and proteins can be induced when the random lateral motion
is mediated by the electrostatic interactions and membrane curvature. Although the lateral diffusion rates of
lipids on membranes of various compositions are measured and the electrostatic free energies of predetermined
protein-membrane-lipid systems can be computed, the process of the aggregation and the evolution to the
electrostatically favorable states remain largely undetermined. Here we propose an electrodiffusion model, based 
on the variational principle of the free energy functional, for the self-consistent lateral drift-diffusion of
multiple species of charged lipids on membrane surfaces. Finite sizes of lipids are modeled to enforce the
geometrical constraint of the lipid concentration on membrane surfaces. A surface finite element method is
developed to appropriate the Laplace-Beltrami operators in the partial differential equations (PDEs) of the
model. Our model properly describes the saturation of lipids on membrane surfaces, and correctly predicts that
the MARCKS peptide can consistently sequester three multivalent phosphatidylinositol 4,5-bisphosphate (\pip2) lipids
through its basic amino acid residues, regardless of a wide range of the percentage of monovalent
phosphatidylserine (PS) in the membrane.
\end{abstract}


\maketitle



\section{Introduction}
Lateral motion of orientated lipids is an intrinsic membrane property. Through specific spatial and temporal variations of their
lateral distribution the lipids regulate many essential membrane-related physiological processes. Multivalent phosphatidylinositol
4,5-bisphosphate (\pip2), for instance, plays a central role in anchoring proteins to plasma membranes and in regulating many ion
channels, enzymes and cytoskeletons \cite{McLaughlinS2002a,SuhB2005a,HoningS2005a,JanmeyP2004a}. Charged lipids are also involved
in driving membrane curvature that is required by cell endocytosis and exocytosis, as illustrated by the lateral movement of
monovalent dioleoylphosphatidylserine (DOPS) to favor electrostatic interactions between negatively charged high-curvature
membranes and positively charged concave surfaces of Bin/amphiphysin/Rvs (BAR)
proteins \cite{BloodP2006,McMahonH2005}. On the other hand, the lateral distribution of lipids are strongly affected by 
the interactions among lipid molecules, binding proteins, membrane curvature and membrane potential, and the ionic distributions 
in the solution \cite{HinderliterA2001a,MclaughlinS2005a,BloodP2006,PetrovA2006a,FilippovA2009a}. It is therefore of great importance to study the
lateral distribution of lipids under a wide range of physiological conditions to provide qualitative or quantitative
information of related physiological processes.

Lateral diffusion is a well recognized model of in-plane lipid motion \cite{FaheyP1977a,TanakaK1999a,AytonG2004a}. Various
techniques such as fluorescence recovery after photobleaching (FRAP), nuclear magnetic resonance (NMR), electron spin
resonance (ESR), and fluorescence spectroscopy have been developed to determine the lateral diffusion coefficients
of lipids \cite{KuoA1979a,EllensJ1993a,CicutaP2007a,ChenY2006a}, mostly by fitting the measured lengths of random walks
to the linear dependence relation $\< r^2(t) \> = D t$ of normal diffusion to determine the
the diffusion coefficient $D$. Surprisingly, the measured coefficients can differ by 2 orders of magnitude
depending on the particular techniques, materials and environments ({\it in vitro} or {\it in vivo}, for example) in
experiments \cite{TocanneJ1989a,VazW1991a,VattulainenI1005a}. Furthermore, diffusion in membranes {\it in vivo} may not follow
the normal linear relation because the membrane contains a number of non-lipid components of different dimensions, which make
the diffusion abnormal. According to the classical free-volume diffusion theory of Cohen and
Turnbull \cite{CohenM1959a,TurnbullD1961a,MaccarthyJ1982a}, lateral diffusion indeed follows a three-step procedure:
(i) creation of a local free volume due to spontaneous thermal fluctuations; (ii) hopping
of a diffusive molecule into the void; and (iii) filling of the hole left behind the hopping molecule
by another molecule. Along with other attempts to describe the lipid diffusion in real complex membrane
systems \cite{SaxtonM1982a}, this theory has been integrated with the scaled-particle theory \cite{LebowitzJ1965a} to model
the dependence of the diffusion coefficient on the sizes of lipids and non-lipid components \cite{OlearyJ1984a}, and the theoretical
results matched the experiments better than the original free-volume diffusion theory. Recent molecular dynamics (MD) simulations
of membrane dynamics have provided different insights of the mechanism of lateral diffusion of lipids. It was shown that the motion
of neighboring lipids are strongly correlated, therefore motion of lipids in clusters is more favored over the hopping of individual
lipids \cite{FalckE2008a}.

Applications of these measured lateral diffusion coefficients to computational simulations or quantitative analysis of
physiological processes, however, are scarce, probably due to a lack of appropriate models of the mechanisms that can
lead to the variation of the lateral distribution of lipids. Suppose that the lateral diffusion of lipids in a
bounded plenary 2-D domain $\Omega$ is described by the standard diffusion equation with source
\begin{equation} \label{eqn:model_0}
\frac{\partial \rho}{\partial t} = D \Delta \rho + f(x,y),
\end{equation}
where $D$ is the constant diffusion coefficient, then the variation of the continuous lipid concentration $\rho$ can be caused by
its initial non-uniform distribution, the product function $f(x,y)$ that models the addition or removal of lipids in the domain,
or the prescribed distribution of $\rho$ on the boundary of $\Omega$. None of these three features have been well modeled for
lipid diffusion. We note, however, that similar models are developed for lateral diffusion of membrane proteins. For instance
the aggregation of proteins to the growing clusters in membranes is modeled as an absorbing boundary condition on the moving
boundary \cite{AmatoreC2009a,AmatoreC2009b}. For
lateral diffusion of lipids or proteins on curved biological membranes one needs to replace the Laplace operator
in Eq.(\ref{eqn:model_0}) by the Laplace-Beltrami operator $\Delta_s$ so the effects of
surface curvatures on the diffusion rate can be incorporated \cite{YoshigakiT2007a,LeitenbergerS2008a,GozdaW2008a}.
Such a model also enables us to investigate the long-term effects of the random fluctuations of the surface metric
on the lateral diffusion \cite{ChevalierC2007a}. More importantly, membrane proteins and many regulatory lipids such
as \pip2 and DOPS are charged hence a drift-diffusion equation is more suitable than the standard diffusion
equation (\ref{eqn:model_0}). Electrodiffusion of charge particles in solution has been extensively studied
in past decades, and there has been resurgent interests recently in modifying the classical
Poisson-Nernst-Planck (PNP) theory to better model the drift-diffusion of charged particles with finite sizes
and particle-particle correlations \cite{Kilic2007b,ZhouY2011a}. PNP theory and its modifications have been extended 
recently to study the lateral motion of lipids or proteins in membranes. For example, 2-D Poisson-Boltzmann-Nernst-Planck 
equations with size exclusion are derived from a constraint free energy minimization to study the lateral motion of lipids 
in the membrane \cite{KiselevV2011a}, assuming the co-existence of mobile ions, peptides, and lipids in the hydrated leaflets
of the membrane. Formation and disassociation of peptide-lipid complexes are modeled via reaction terms in 
the Nernst-Planck equations, and results are compared favorably to their dynamic Monte Carlo 
simulations \cite{KiselevV2011b}. 2-D PNP equations without size exclusion were used to simulate the electrodiffusion of 
lipids on a planar surface \cite{KhelashviliG2008a}, coupled with a nonlinear 3-D Poisson-Boltzmann equation with a 
charge density neglecting the contribution of charged lipids. While various degrees of agreement to the experimental 
measurements or other theoretical predictions have been achieved with these models, an electrodiffusion model similar to 
the modified Poisson-Nernst-Planck equations in bulk are necessary for quantitative study of self-consistent distributions of 
charged particles on general curved surfaces.

In this paper we developed a generic surface drift-diffusion equation based on a generalized Borukhov
model \cite{Borukhov1997,ZhouY2011a} to model the electrostatic mediated lateral motion of charged lipids on arbitrarily 
curved membrane surfaces. The bilayer membrane is described as a dielectric continuum with a constant permittivity and 
continuous distributions of surface charges, the latter model the distributions of charged lipids on membrane surfaces. 
The lipids are treated as diffusive hard disks, and their effective radius are taken into account together with the finite 
sizes of mobile ions in the entropic contribution of the total free energy. Without proper modeling of the lipid sizes the 
lateral concentration of charged lipids can be easily overestimated wherever there are strong attractive electrostatic
interactions, even though the molecular dimensions of lipids have been considered in the determination of the lateral
diffusion coefficient. We derive a system of nonlinear partial differential equations (PDEs) and develop the related
numerical algorithms. Numerical simulations are carried out to study the surface electrodiffusion of \pip2 due to
the electrostatic interactions between the membrane and the MARCKS protein under a wide range of physiological conditions.
Our results are well validated by experimental measurements and atomistic modeling of lipid sequestration.

\section{Mathematical Models of Surface Electrodiffusion}
\begin{figure}[!ht]
\begin{center}
\includegraphics[width=1.6in]{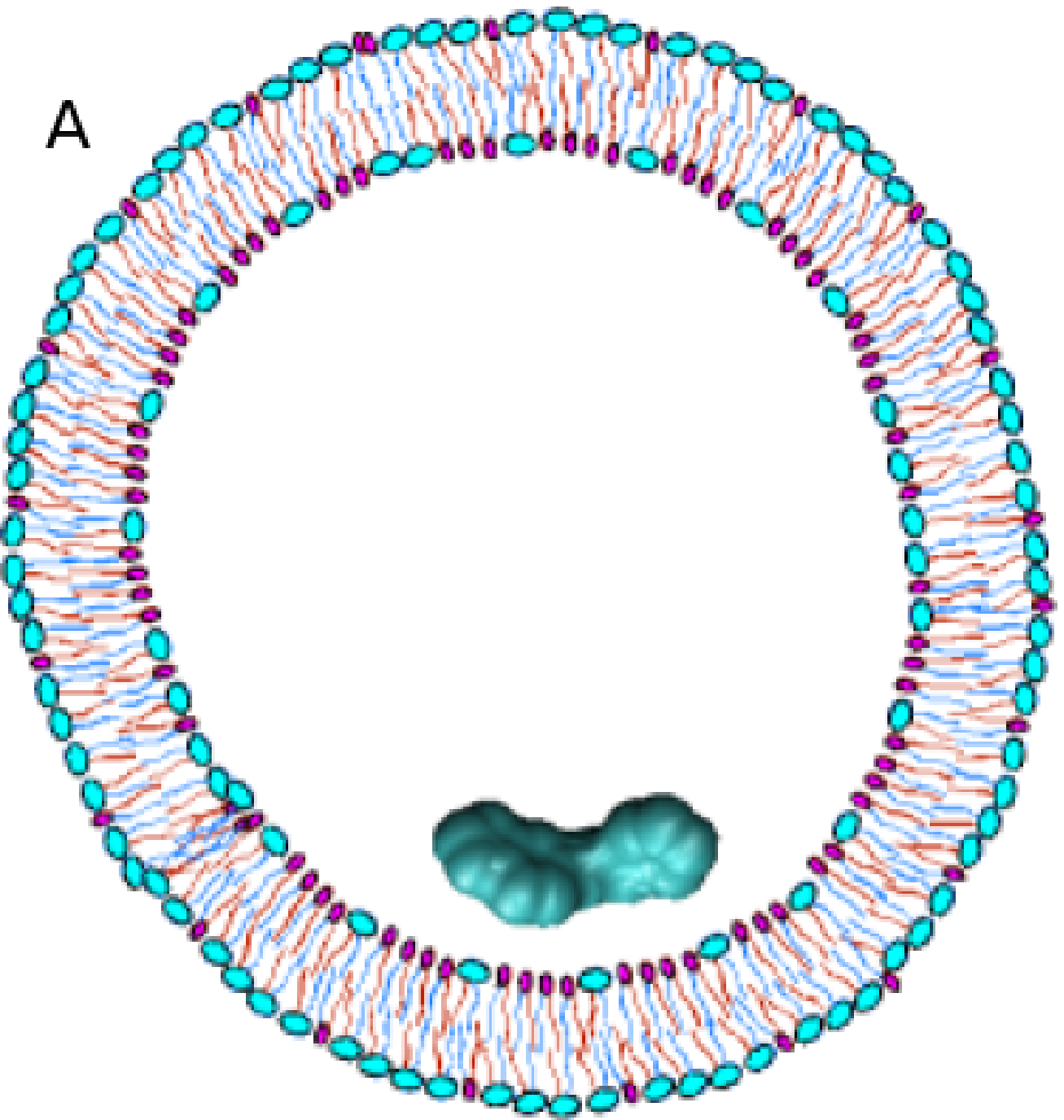}
\includegraphics[width=1.6in]{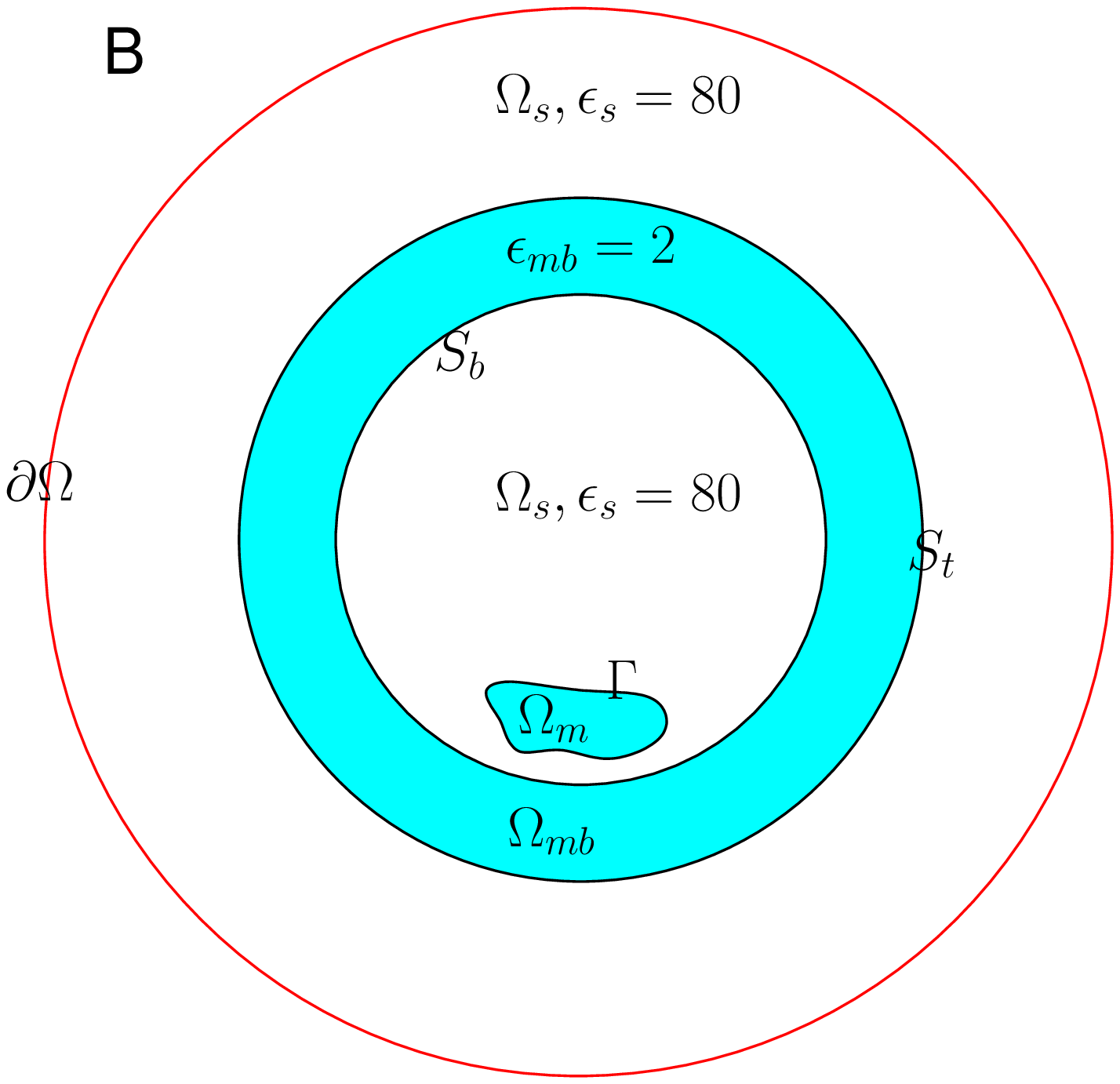}
\includegraphics[width=1.6in]{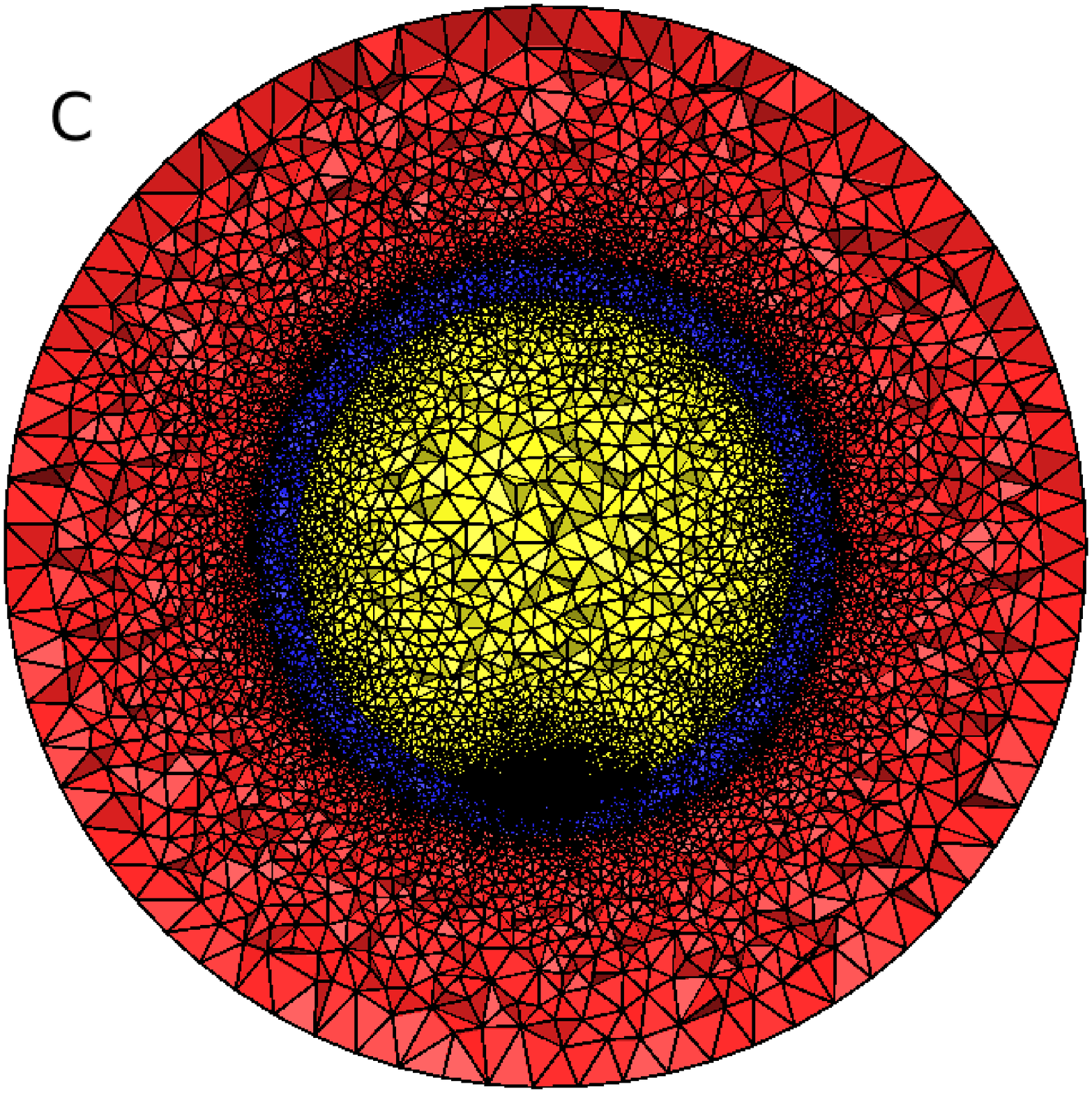}
\includegraphics[width=1.6in]{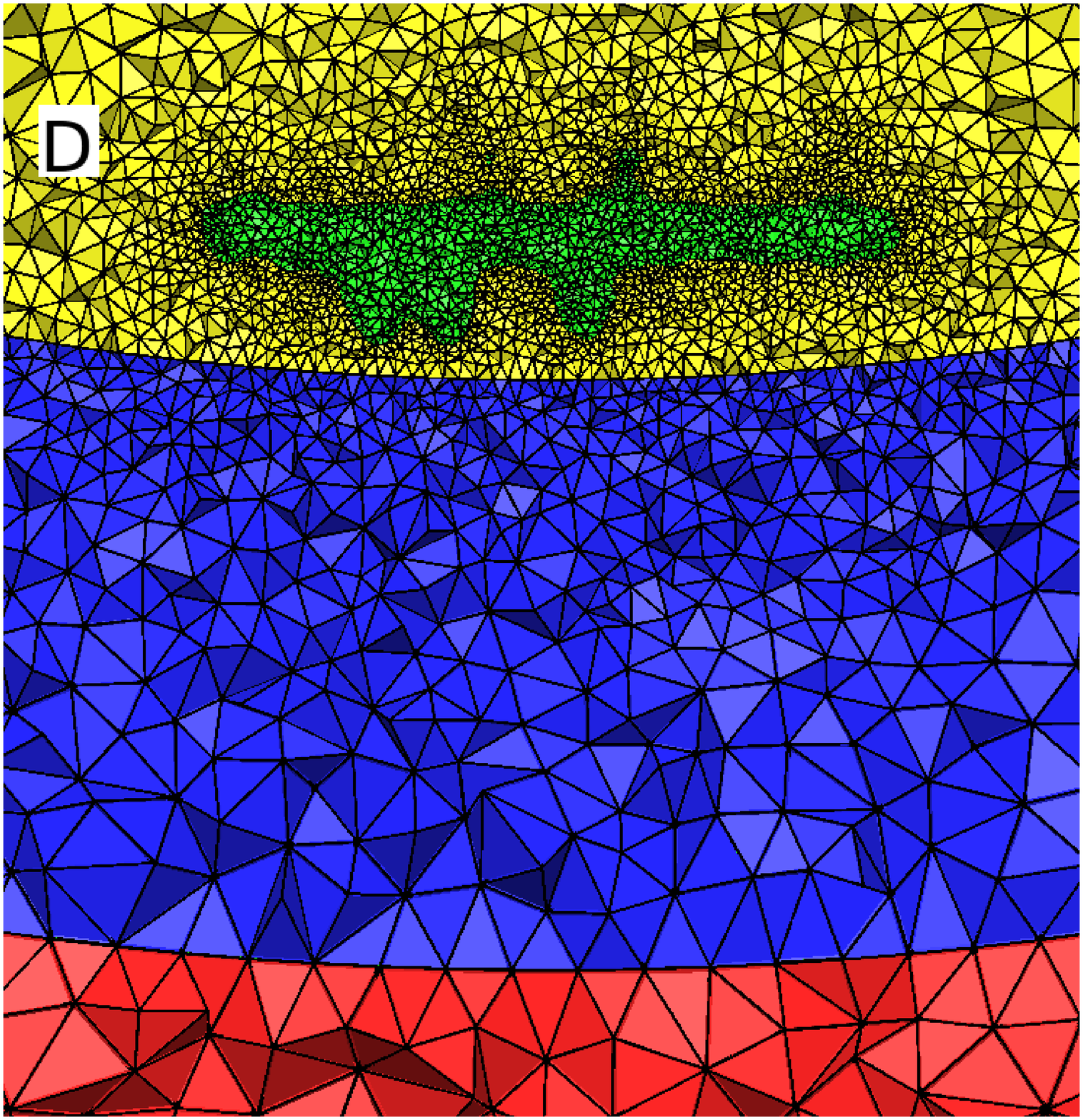}
\end{center}
\caption{
{\bf Solvated MARCKS-membrane system and the computational model.} (A) Schematic illustration of the MARCKS peptide residing on
the inner leaflet of a vesicle. Lipids with charged (red) and polar (cyan) headgroups are distributed in the membrane.
(B) 2D cross section of the computational model for (A). The domain $\Omega_{mb}$ is the bilayer membrane modeled as a
dielectric continuum without atomistic details. Distribution of lipids on membrane surfaces $S_t$ and $S_b$ follows the surface
electrodiffusion equation. Mobile ions are distributed in the aqueous solution exterior and interior to the vesicle. Atomistic
structure of the MARCKS peptide is retained and singular charges are distributed in $\Omega_m$. $\Gamma$ is the molecuar surface
of the peptide, and $\partial \Omega$ is the exterior boundary of the computational domain for the Poisson equation. (C) 3D 
tetrahedral discretization of the entire computational domain. (D) local refinement of the tetrahedral mesh near the 
MARCKS peptide.}
\label{fig:system}
\end{figure}

We consider a solvated protein-membrane system as illustrated in Figure \ref{fig:system}(A). In the spirit of the
classical fluid density functional theory \cite{Burak01,ZhouY2011a,KiselevV2011a} we define the following free energy 
for the membrane-protein-ionic solution system:
\begin{widetext}
\begin{eqnarray}
F & = & k_B T \int_{\Omega_s} \sum_{i=0} \rho_i  \left[ \ln \left( \frac{\rho_i}{\xi_i} \right) - 1 \right] dx +
    k_B T \int_{S_t \cup S_b}  \sum_{j=0} \rho^l_j \left[ \ln \left( \frac{\rho^l_j}{\xi^l_j} \right) - 1 \right] ds 
   + \nonumber \\
  &  &  \int_{\Omega} -\frac{1}{2} \epsilon | \nabla \phi |^2 +
\left( \rho^f + \lambda \sum_{i=1} z_i e \rho_i + \sum_{j=1} z_j^l e \rho^l_j
\delta(x - X_{S_t \cup S_b}) \right ) \phi dx  \label{eqn:energy}
\end{eqnarray}
\end{widetext}
where $\rho_i$ is the ion concentration of the $i^{th}$ species and $\rho^l_j$ is the lipid concentration of $j^{th}$ species,
with $i=0$ denotes the neutral solvent (water) molecules and $j=0$ denotes the neutral species of lipids. The characteristic 
function $\lambda$ is $1$ in the solution domain $\Omega_s$ and zero elsewhere. $X$ is the coordinates of the membrane 
surfaces $S_t$ and $S_b$. $k_B,T,\phi$ are the Boltzmann constant, temperature, and the electrostatic potential,
respectively. $z_i, z_j^l$ are the valences of corresponding ions and lipids, and $e$ is the elementary charge.
$\rho^f$ is the permanent charges in proteins.
$$\xi_i = \frac{1}{a_i^3}$$
is the fugacity of the $i^{th}$ species of ions with effective size $a_i$, and
$$\xi_j = \frac{1}{(a^l_j)^2}$$
is the fugacity of $j^{th}$ species of lipids approximated as hard disks with an effective size $a^l_j$ 
\cite{theoryofsimpleliquids,ZhouY2011a}. Here the solvent molecules and neutral lipids have respective effective 
sizes $a_0$ and $a_0^l$, and their concentrations $\rho_0, \rho_0^l$ follow the relations
\begin{eqnarray}
\rho_0 a_0^3 + \sum_{i=1} \rho_i a_i^3 & =  & 1, \label{eqn:water} \\
\rho^l_0 (a^l_0)^2 + \sum_{j=1} \rho^l_j (a^l_j)^2 & =  & 1. \label{eqn:neutral_lipids}
\end{eqnarray}
These two constraints are enforced directly in the variations of $F$ with respect to $\rho_i$ and $\rho_i^l$
without resorting to a Lagrangian multiplier, in contrast to \cite{KiselevV2011a}, 
giving rise to the electrochemical potentials $\mu_i$ for ions ($i \ge 1$) and $\mu^l_j$ for charged lipids ($j \ge 1$):
\begin{widetext}
\begin{eqnarray}
\mu_i  & = & \frac{\delta F}{\delta \rho_i} = z_i e \phi +
                        k_B T \left[ \ln (\rho_i a_i^3) - k_i \ln \left( 1 - \sum_{k=1} \rho_k a_k^3  \right)  \right],  \label{eqn:mu_ion} \\
\mu^l_j& = & \frac{\delta F}{\delta \rho^l_j} = z_j^l e \phi +
                        k_B T \left[ \ln (\rho_j^l (a^l_j)^2) - k^l_j \ln \left( 1 - \sum_{k=1} \rho_k^l (a^l_k)^2 \right) \right], \label{eqn:mu_lipid}
\end{eqnarray}
\end{widetext}
where $k_i = (a_i/a_0)^3, k^l_j = (a^l_j/a^l_0)^2$. 
These electrochemical potentials are related to the flux of ions and lipids through the constitutive relations for fluxes
\begin{equation*}
J_i =-m_i \rho_i \nabla \mu_i,  \quad  
J^l_j  = -m^l_j \rho^l_j \nabla_s \mu^l_j, 
\end{equation*}
where $\nabla_s$ is the surface gradient operator. $m_i$ and $m^l_j$ are the mobilities of respective species of ions
and lipids; they are related to the diffusivities $D_i$ or $D^l_j$ through Einstein's relation
$$ D_i = m_i k_B T \quad \mbox{or} \quad D^l_j = m^l_j k_B T.$$
The drift-diffusion equations for ions are obtained from the mass conservation in solution:
\begin{align}
& \frac{\partial \rho_i}{\partial t} = -\nabla \cdot J_i = \nabla \cdot (m_i \rho_i \nabla \mu_i) \nonumber \\
& = \nabla \cdot D_i \left ( \nabla \rho_i + \frac{k_i \rho_i \displaystyle{ \sum_{p=1} a_p^3 \nabla \rho_p}}{1 -
\displaystyle{ \sum_{p=1} a_p^3 \rho_p} } + \frac{1}{k_B T} \rho_i z_i e \nabla \phi \right),  \label{eqn:NP_ion}
\end{align}
and the surface drift-diffusion equations for lipids are obtained from the mass conservation on the membrane surfaces if it evolves
at a given velocity field $u_s$ \cite{StoneH1990a}:
\begin{widetext}
\begin{align}
\frac{\partial \rho^l_j}{\partial t} + \nabla_s \cdot (\rho^l_j u_s) + \rho^l_j (\nabla_s \cdot n) (u_s \cdot n) & 
= -\nabla_s \cdot J^l_j = \nabla_s \cdot (m^l_j \rho^l_j \nabla_s \mu^l_j) \nonumber \\
& = \nabla_s \cdot D_j^l \left ( \nabla_s \rho^l_j + \frac{k^l_j \rho^l_j \displaystyle{ \sum_{p=1} (a^l_p)^2 \nabla_s \rho^l_p}}{1 -
\displaystyle{ \sum_{p=1} (a^l_p)^2 \rho^l_p} } + \frac{1}{k_B T} \rho^l_j z^l_j e \nabla_s \phi \right), \label{eqn:NP_lipid}
\end{align}
\end{widetext}
where $n$ is the outer normal vector of the surface. If there are only two types of lipids of the same effective size,
one charged and the other neutral, in a macroscopically static membrane, one can solve only one surface drift-diffusion equation,
which is now simplified to be
\begin{align}
\frac{\partial \rho^l}{\partial t} & = \nabla_s \cdot D^l \left ( \nabla_s \rho^l + \frac{k^l (a^l)^2 \rho^l \nabla_s \rho^l}{1 -
(a^l)^2 \rho_p} + \frac{1}{k_B T} \rho^l z^l e \nabla_s \phi \right), \label{eqn:NP_lipid_final}
\end{align}
in which the subscription $j$ for species is no longer needed. For enclosed membrane surfaces Equations (\ref{eqn:NP_lipid})
and (\ref{eqn:NP_lipid_final}) do not have boundary conditions, but both are subject to the constraint of mass conservation
\begin{equation}
\int_{S} \rho_j^l ds = T_j, \label{eqn:rho_total}
\end{equation}
where $T_j$ is the given total quantity of a species of charged lipids on the surface $S$. Extremization of $F$ with respect
to $\phi$ using variation $\delta F/\delta \phi$ gives the Poisson equation for the electrostatic potential $\phi$:
\begin{equation} \label{eqn:Poisson}
- \nabla \cdot (\epsilon \nabla \phi) = \rho^f + \lambda \sum_{i=1} \rho_i z_i e \quad \mbox{in}~ \Omega,  
\end{equation}
with interface conditions
\begin{eqnarray*}
\epsilon_{mb} \frac{\partial \phi_{mb}}{\partial n} = \epsilon_s \frac{\partial \phi_{s}}{\partial n} + \sum_{j=1} \rho_j^l z_j^l e, 
\quad \phi_s = \phi_{mb} ~ \mbox{on}~ S_t \cup S_b, \\
\epsilon_{m} \frac{\partial \phi_{m}}{\partial n} = \epsilon_s \frac{\partial \phi_{s}}{\partial n}  \quad \phi_s = \phi_{m}
~ \mbox{on}~ \Gamma,
\end{eqnarray*}
where the charge density carried by charged lipids on the membrane surfaces is written as an interface condition. 
The subscript $m,mb,s$ denote the quantities in proteins, membrane, and the solution, respectively.

To acquire an accurate characterization of the diffusion process we will solve the Poisson equation (\ref{eqn:Poisson}) 
coupled with the time-dependent drift-diffusion equation (\ref{eqn:NP_ion}) and the time-dependent surface drift-diffusion
equation (\ref{eqn:NP_lipid}), using a finite element method \cite{ZhouY2011a} and a linear surface finite element 
method \cite{Dziukg1988a}. If only the equilibrium distributions of ions and charged lipids are of interest one can 
instead solve the Poisson equation
coupled with steady-state drift-diffusion equation and stead-state surface electrodiffusion equation. Notice the latter,
\begin{widetext}
\begin{equation}
\nabla_s \cdot D_j^l \left ( \nabla_s \rho^l_j + \frac{k^l_j \rho^l_j \displaystyle{ \sum_{p=1} (a^l_p)^2 \nabla_s \rho^l_p}}{1 -
\displaystyle{ \sum_{p=1} (a^l_p)^2 \rho^l_p} } + \frac{1}{k_B T} \rho^l_j z^l_j e \nabla_s \phi \right) = 0,
\label{eqn:NP_lipid_steady}
\end{equation}
\end{widetext}
admits a trivial solution $\rho_j^l=0$ if the constraint (\ref{eqn:rho_total}) is not enforced. To obtain a physical solution that is
consistent with this constraint we introduce the decomposition
\begin{equation}
\rho_j^l = \bar{\rho}_j^l + \hat{\rho}_j^l,
\end{equation}
where
\begin{equation}
\bar{\rho}_j^l = \frac{T_j}{|S|}
\end{equation}
is the known average concentration of $\rho_j^l$ on the surface $S$ with total surface area $|S|$.
Replacing $\rho_j^l$ in equation (\ref{eqn:NP_lipid_steady}) with this decomposition we get
\begin{widetext}
\begin{equation}
\nabla_s \cdot D_j^l \left ( \nabla_s \hat{\rho}^l_j +
\frac{k^l_j (\hat{\rho}^l_j + \bar{\rho}^l_p) \displaystyle{ \sum_{p=1} (a^l_p)^2 \nabla_s \hat{\rho}^l_p}}{1 -
\displaystyle{ \sum_{p=1} (a^l_p)^2 (\hat{\rho}^l_p + \bar{\rho}^l_p ) } } +
\frac{1}{k_B T} \hat{\rho}^l_j z^l_j e \nabla_s \phi \right) =
-\nabla_s \cdot \left ( D^l  \frac{1}{k_B T} \bar{\rho}^l_j z^l_j e \nabla_s \phi \right ),
\label{eqn:NP_lipid_steady_var}
\end{equation}
\end{widetext}
from  which the nontrivial variation $\hat{\rho}_j^l$ and hence the nontrivial $\rho_j^l$ can be uniquely solved.

\section*{Solvated MARCKS-membrane system}

The system, as illustrated by Figure \ref{fig:system}(A-B), consists of a MARCKS peptide and a membrane vesicle whose interior and
exterior radii are $230$\AA~and $270$\AA, respectively. The MARCKS peptide is unfolded and the peptide is built from the sequence of the
effective domain as given in \cite{ArbuzovaA2002a}. The peptide is placed above the inner leaflet of
the vesicle, the closest distance between the peptide and the membrane surface is about $2$\AA, with the assumption that
the position and orientation are not affected by its interaction with the membrane and lipids. The insertion of MARCKS is not
modeled here because the aromatic phenylalanine
residues penetrating to the acyl chain region obstruct the diffusion of lipids. Experiments and Monte Carlo simulations
suggest that the diffusion coefficients measured on systems free of inserted molecules may not be applicable
to the obstructed diffusion \cite{RattoT2002a,ShortenP2009a}. In order to investigate electrodiffusion of lipids with
different valences we will consider two types of lipid compositions, PIP2/PC(Phosphatidylcholine) and PIP2/PS(phosphatidylserine)/PC;
all head groups are assumed to have the same effective diameter $8.3$\AA. The two leaflets of the membrane are assigned the same average
lipid composition. The dielectric constant $\epsilon_m = \epsilon_{mb} = 2$ in the peptide and the membrane, and $\epsilon_s=80$ in
the solution. There are significant disparities in the measured lateral diffusion coefficients of lipids in the plasma membrane
\cite{EllensJ1993a,FlennerE2009a}, here we choose $D^l = 580$ \AA$^2/\mu s$. The solution contains the salt KCl only, whose
concentration will be adjusted between 50mM to 300mM for the examination of the effects of the salt concentration on the lateral
electrodiffusion. The effective sizes of K and Cl are $2.7$\AA~and $3.6$\AA, respectively, and their diffusion coefficients
are $78000$\AA$^2/\mu$. We note that the final equilibrium distributions of lipids and mobile ions are independent of these
diffusion coefficients. There is no ion exchange through the membrane. The Poisson equation (\ref{eqn:Poisson}) 
is solved in the entire 3-D domain $\overline{\Omega_s \cup \Omega_{mb} \cup \Omega_m}$, 
the bulk drift-diffusion equation (\ref{eqn:NP_ion}) is solved in the 3-D solvent domain $\Omega_s$, and the
surface drift-diffusion equation is solved on both spherical membrane surfaces $S_t$ and $S_b$. This way the lateral 
diffusion of lipids on two membrane leaflets are coupled. A cross section of the finite element tetrahedral mesh
and the locally refined mesh near the peptide are shown in Figure \ref{fig:system}(C-D).

\section{Computational Simulations and Discussions}
\subsection{Effects of finite size of \pip2 on its surface distribution}
\begin{figure}[!ht]
\begin{center}
\includegraphics[width=2in]{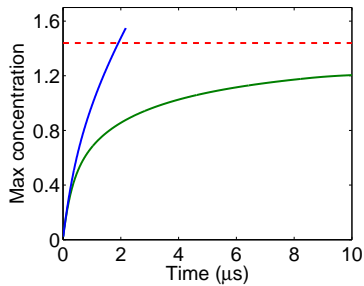}
\end{center}
\caption{
{\bf Without an appropriate model of the lipid size the \pip2 quickly exceeds the upper limit
of the concentration under the attraction of the MARCKS peptide}. Blue:
lipids with vanishing size; Green: lipids with an effective diameter $8.33$\AA.
The red line signifies the limit of the lipid concentration $0.0144/$\AA$^2$.
Initial average percentage of \pip2 is $1$\%. The scale on the vertical axis is
amplified 100-fold.
}
\label{fig:sizeComp}
\end{figure}

We compare the aggregation of two models of \pip2 lipids due to the electrostatic attraction of the MARCKS peptide.
In one model the lipids are described as particles with vanishing sizes and in the other the lipids have an
effective diameter $8.33$\AA. We observe the aggregation by starting with an initial uniform distribution of 1\% \pip2
on membrane surfaces. The histories of aggregation are shown in Figure \ref{fig:sizeComp}. The accumulation of charged
lipids is found much faster when the size effects are neglected, and its concentration will quickly exceed the upper
limit of the lipid concentration, resulting in an unphysical distribution
of lipids on membrane surfaces. This upper limit of the concentration of \pip2 (i.e., the membrane consisting
of 100\% \pip2) is $0.0144$/\AA$^2$, corresponding to about 36 lipids on a 50$\times$50\AA$^2$
membrane surface \cite{BloodP2006}. The overestimated concentrations of charged lipids can
lead to an overestimated dielectric surface force density in studying the protein-membrane electrostatic
interactions \cite{ZhouY2010b}. In contrast the model with lipid size correction predicts a lipid
concentration that is well bounded by the saturation value.

\subsection{Dependence of \pip2 aggregation on its average concentration}
\begin{figure}[!ht]
\begin{center}
\includegraphics[width=2.0in]{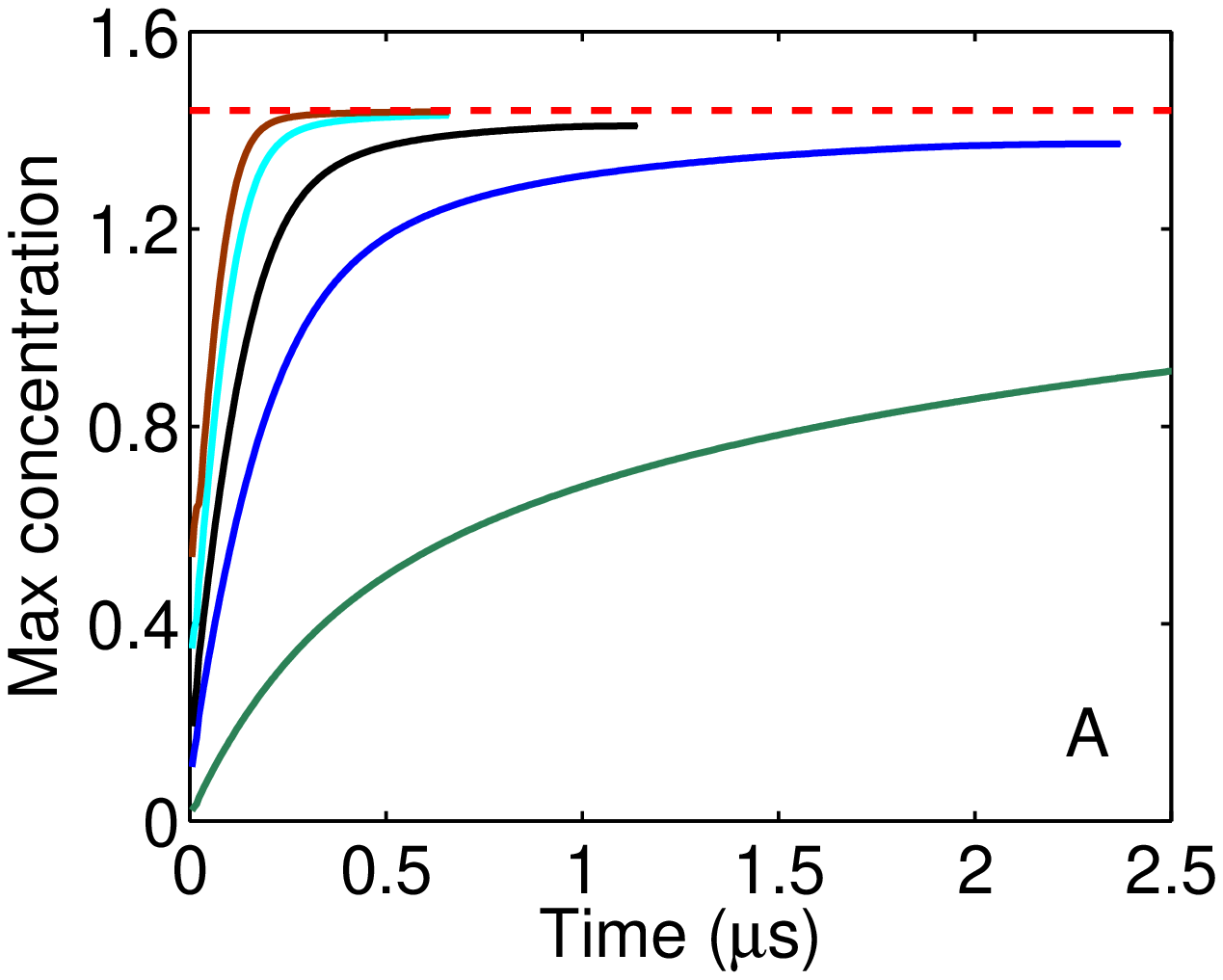}
\includegraphics[width=2.0in]{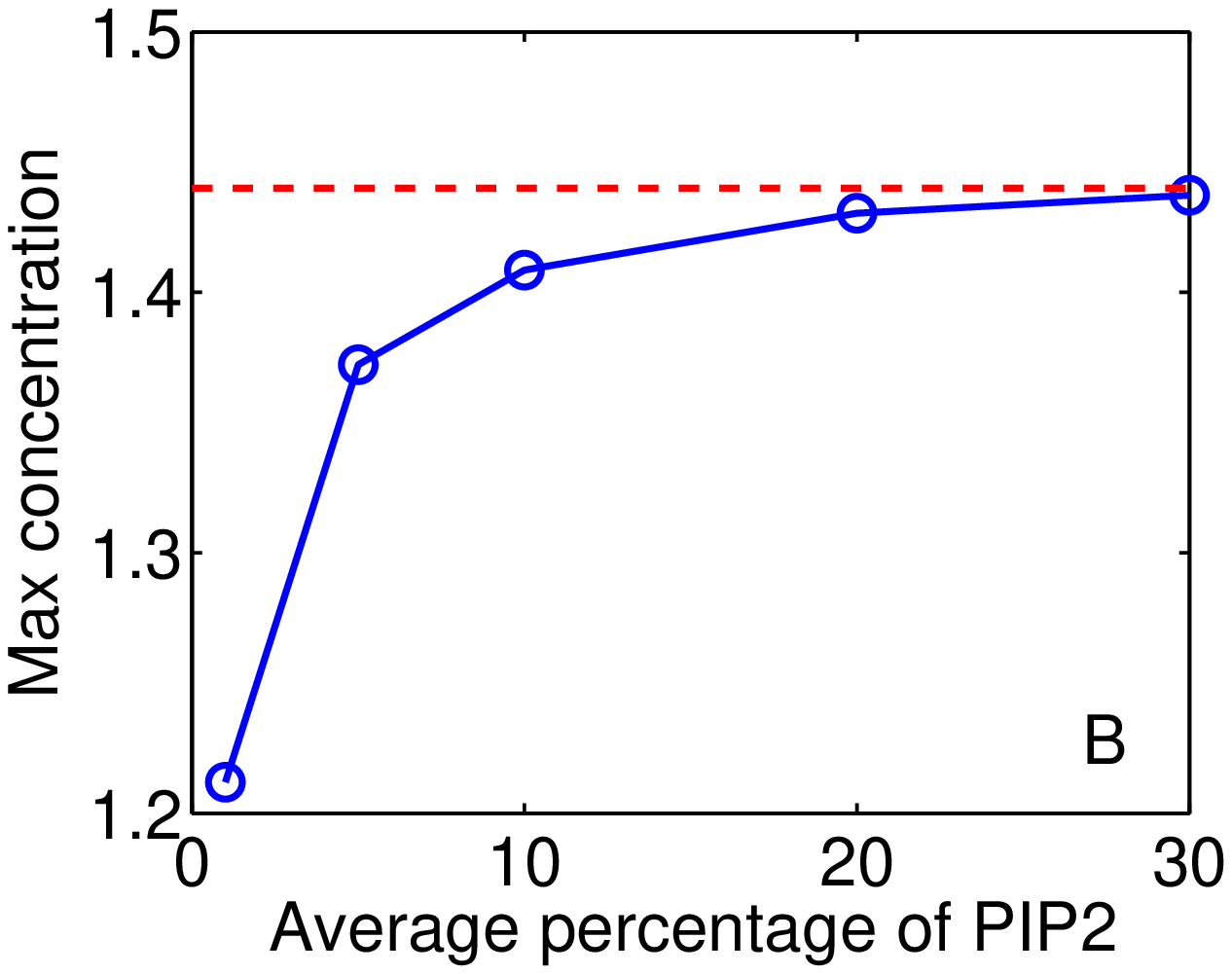}
\includegraphics[width=2.0in]{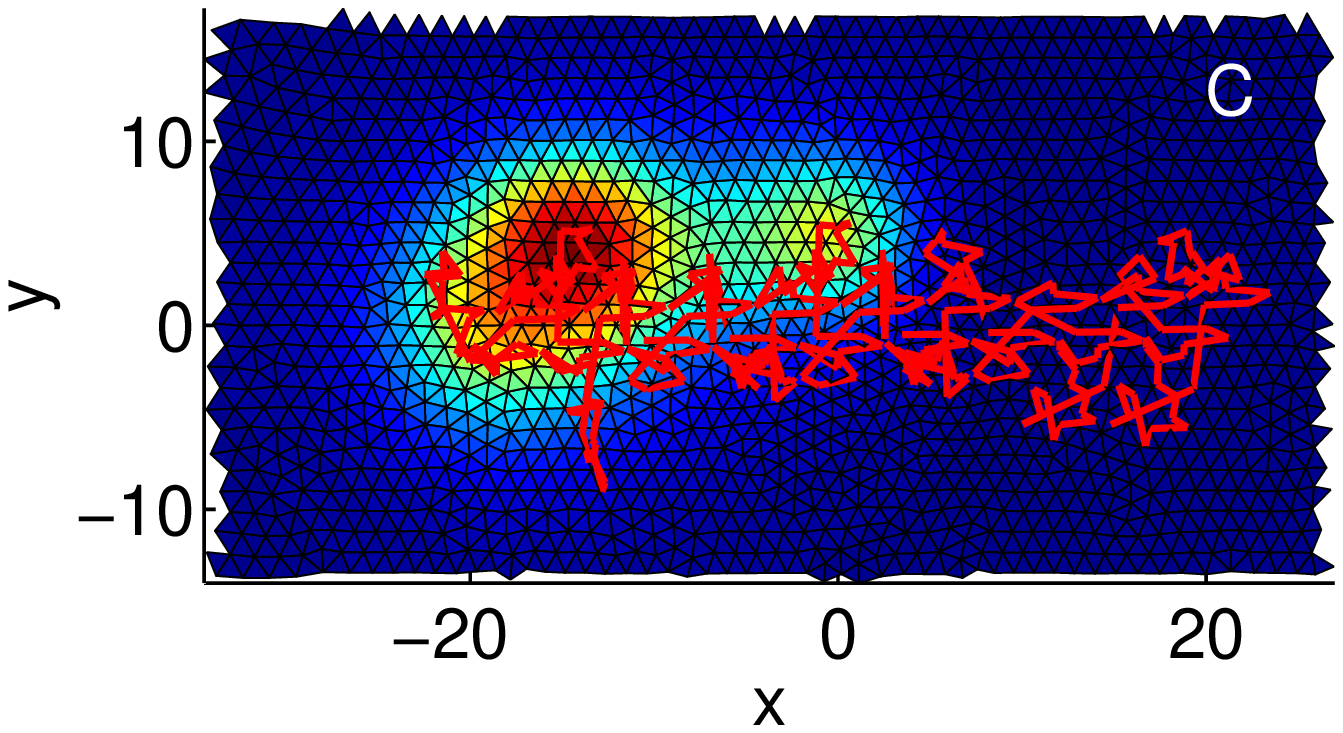}
\includegraphics[width=2.0in]{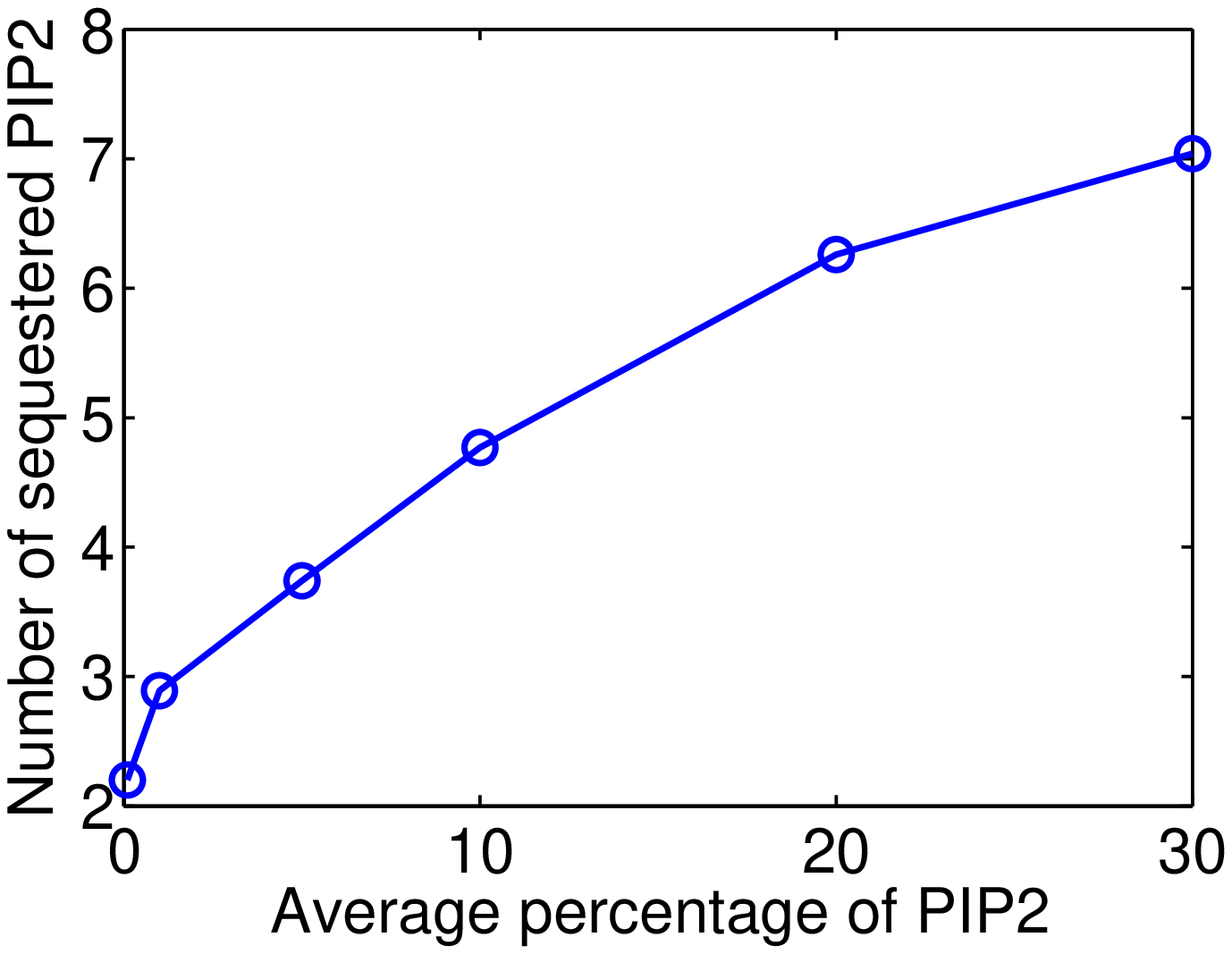}
\end{center}
\caption{
{\bf (A) Accumulation of \pip2 near the MARCKS peptide; (B) the dependence of the maximum concentration
of \pip2 on the initial average concentration; (C) Integration region used for counting the sequestered
\pip2; and (D) Number of sequestered \pip2}. The initial average concentrations of \pip2
in (A) are $1$\%, $5$\%, $10$\%, $20$\%, $30$\% for the green, blue, black, cyan, and maroon lines,
respectively. Complete history for the simulation with initial $1$\% \pip2 is shown in Figure
\ref{fig:sizeComp}. In (C) the red lines represents the MARCKS peptide above the membrane surface.
The integration region is not pre-determined for a conforming triangulation so its boundary is not
straight. The average percentages in (D) are 0.1\%, 1\%, 5\%, 10\%, 20\% and 30\%.
}
\label{fig:sionComp}
\end{figure}

Experiments \cite{McLaughlinS2002a,SuhB2005a,JanmeyP2004a} show that a MARCKS peptide can consistently
sequester three or four \pip2 lipids when the average fraction of \pip2 in the membrane varies between 0.01\% and
1\%. Suppose 1\% \pip2 is uniformly distributed in the membrane, 3 \pip2 lipids shall occupy a surface area
of about 144$\times$144\AA$^2$. Such an area is nearly 33 times larger than the area coverage of MARCKS peptide
($\sim$45$\times$15\AA$^2$). For 0.1\% \pip2 this area amounts to almost 1/3 of the total area of the inner leaflet
of the vesicle with an inner radius $230$\AA; and there is barely one \pip2 lipid on the entire inner surface of the
vesicle when the membrane contains only 0.01\% \pip2. Consequently we define the lower bound of the \pip2 percentage in our
simulations to be 0.1\%. The maximum percentage of \pip2 in our simulations is 30\%, for which 3 \pip2 lipids occupy an area of
26.4$\times$26.4\AA$^2$ when distributed uniformly; this area approximates the coverage of MARCKS peptide on the membrane surface.
This general assessment of the sparsity of \pip2 lipids suggests that the equilibration of \pip2 distribution in response to the
electrostatic perturbation of MARCKS peptide will vary as the initial percentage of \pip2 changes. To quantify these differences and to
reveal the continuous dependence of the sequestration of \pip2 on its average concentration we simulate
the electrodiffusion of \pip2 with five initial conditions; the results are summarized in
Figure \ref{fig:sionComp}. With 30\% \pip2 it takes about 0.2$\mu s$ to re-equilibrate its distribution in
response to the approaching of MARCKS peptide, while it takes about 10$\mu s$ if the membrane contains 1\% \pip2.
For an initial percentage of 0.1\% this time becomes as long as 100 $\mu s$. At a low percentage of \pip2 it takes a much
longer time for the lipids to move over a larger distance to the vicinity of the MARCKS peptide so the lipids can be sequestered.
Cautions must be taken in interpreting this long time of aggregating because the attraction of \pip2 to the peptide is not always
started with a uniform distribution of \pip2. Indeed, the MARCKS peptides are recycled in real biological systems: when [Ca$^{2+}$]
increase in the local cytoplasm the Ca$^{2+}$/calmodulin (Ca/CaM) will bind to the MARCKS peptides that have attached to the
membrane and pull them off. When the detached MARCKS proteins move away from the membrane the sequestered \pip2 lipids will be
released and diffuse away from the aggregated region \cite{MclaughlinS2005a}. If the period of this recycle is sufficiently
small, the re-attached MARCKS peptide will quickly attract the \pip2 lipids back before the re-establishment of an
uniform distribution of \pip2. It is therefore possible to integrate the current continuum surface electrodiffusion
equations for lipids with the diffusion-reaction dynamical models for Ca$^{2+}$, calmodulins, and other associated
molecules to yield an accurate characterization of the biochemical signals that are transmitted through these processes.

We count the number of sequestered \pip2 lipids by integrating the density on the membrane surface
$$ N_{\text{\pip2}} = \int_{S_{p}} \rho^l ds,$$
where the sequestration domain $S_{p}$ is defined to be the projection to the inner membrane surface of a
45$\times$15\AA$^2$ rectangular centered at Ser162:N, as shown in Figure \ref{fig:sionComp}(C).
This rectangular is chosen to enclose the entire
MARCKS peptide. We note that this definition of sequestration is different from those in the atomistic
modeling of lipids \cite{McLaughlinS2002a,WangJ2002a}, where \pip2 lipids can be represented as potential surfaces of -25mV. The
sequestration of lipids is identified as the enclosure of these negative potential surfaces by the
iso-surfaces of positive potential ($\sim$25mV) that are induced by the basic residues in the peptide. Quantitative agreements
are found between the current continuum and the previous atomistic models of lipids. At 0.1\% initial percentage,
about 2 lipids will finally be attracted to the membrane surface under the peptide after a long travel, and about 3 \pip2
lipids can be sequestered at 1\% initial percentage.
More \pip2 lipids are found to be clustered below
the peptide with a further increase of its percentage, gradually saturating to the maximum density of
lipids. Figure \ref{fig:sionComp}(B) indicates that the maximum concentration of \pip2 reaches the
saturation density at 30\% of average concentration. Correspondingly there are about 7 \pip2 are
clustered below the peptide, c.f. Figure \ref{fig:sionComp}(D), which account for 72\% of the total
lipids in the domain $S_{p}$. It is worth noting that this aggregated distribution of \pip2 is generated from
the minimization of the total free energy (\ref{eqn:energy}), and is consistent with the electrostatic potential
field of the membrane-MARCKS system by construction. In contrast, with atomistic models of the sequestration one
needs to place charged lipids at positions that are determined {\it a priori} so that the electrostatic potential
can show favorable electrostatic interactions between basic peptides and phospholipid membranes \cite{WangJ2004a}.
The current continuum model thus has a great potential in modeling the membrane-protein systems for which the
positions of specific lipids are unknown {\it a priori} or change dynamically with the conformational change of
membrane or proteins during their interactions.


\subsection{Dependence of \pip2 aggregation on the concentration of monovalent lipids}
\begin{figure}[!ht]
\begin{center}
\includegraphics[width=1.6in]{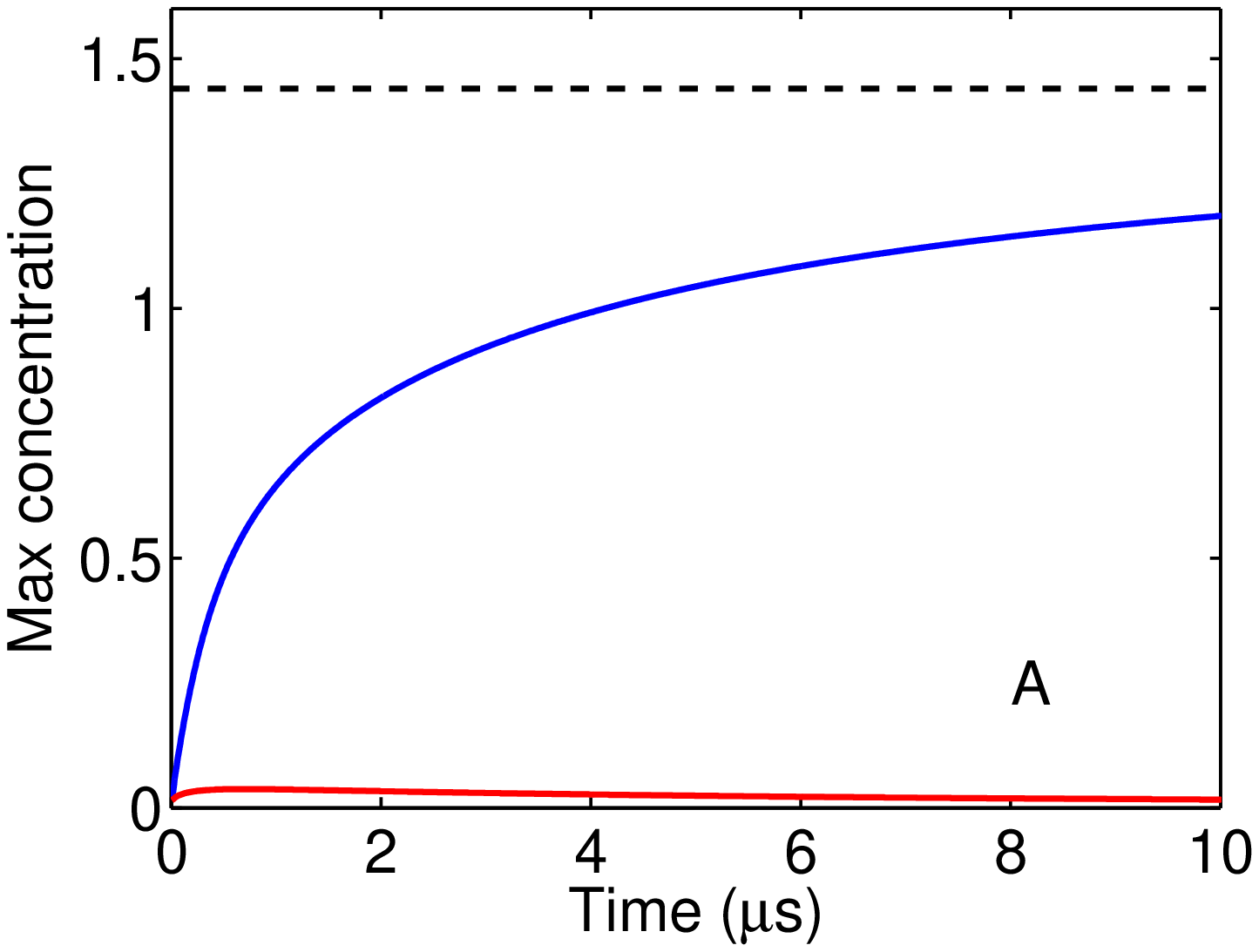}
\includegraphics[width=1.6in]{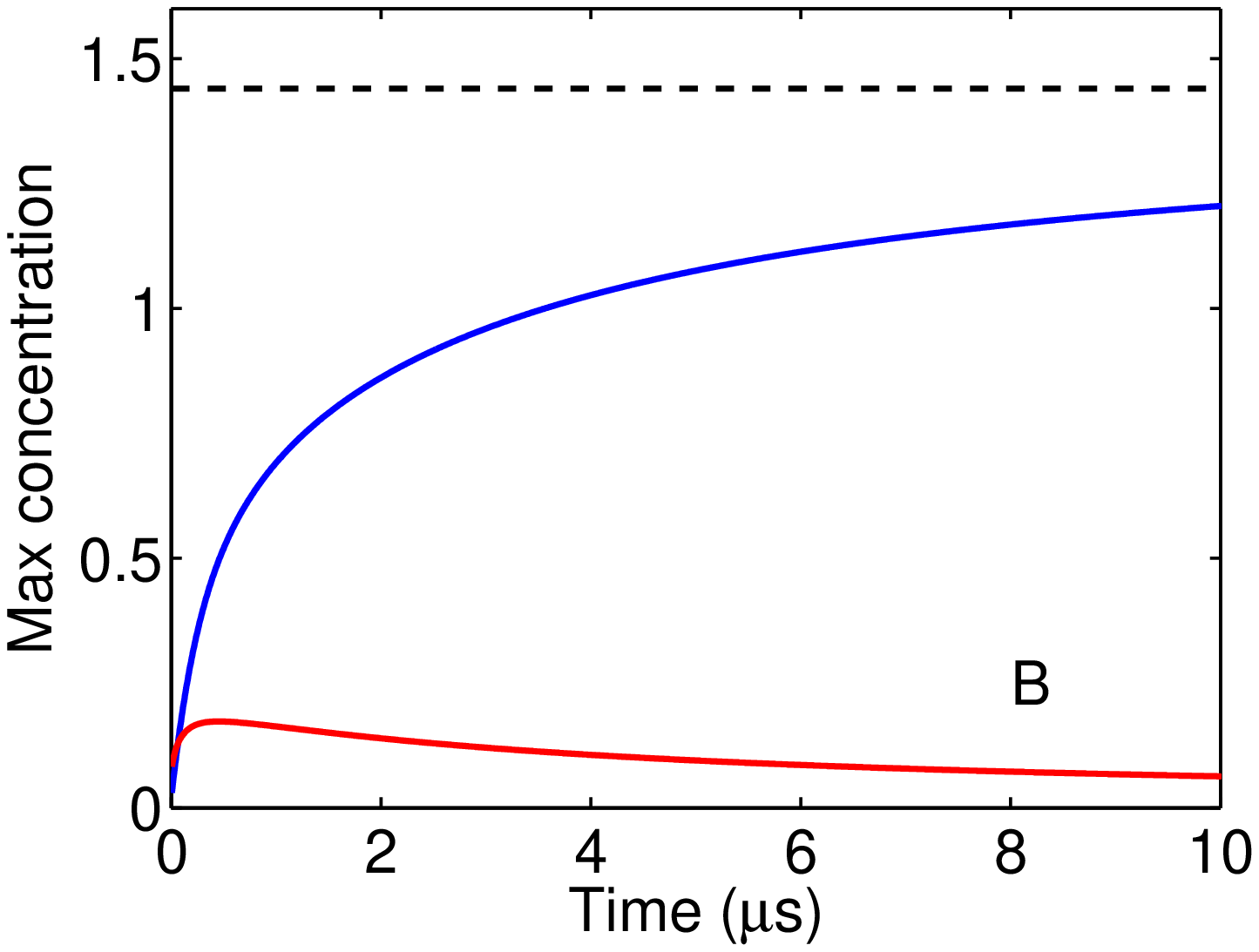}
\includegraphics[width=1.6in]{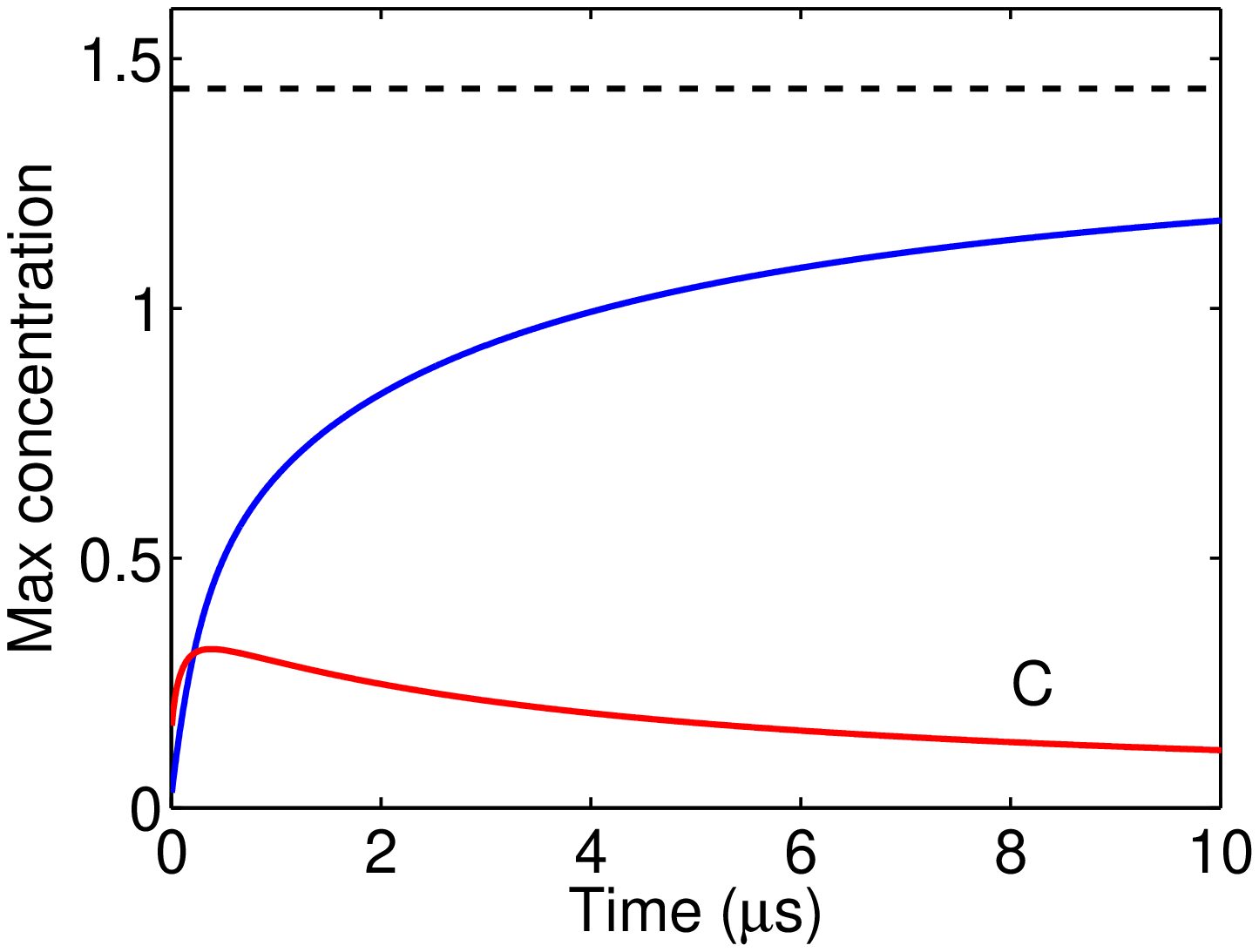}
\includegraphics[width=1.6in]{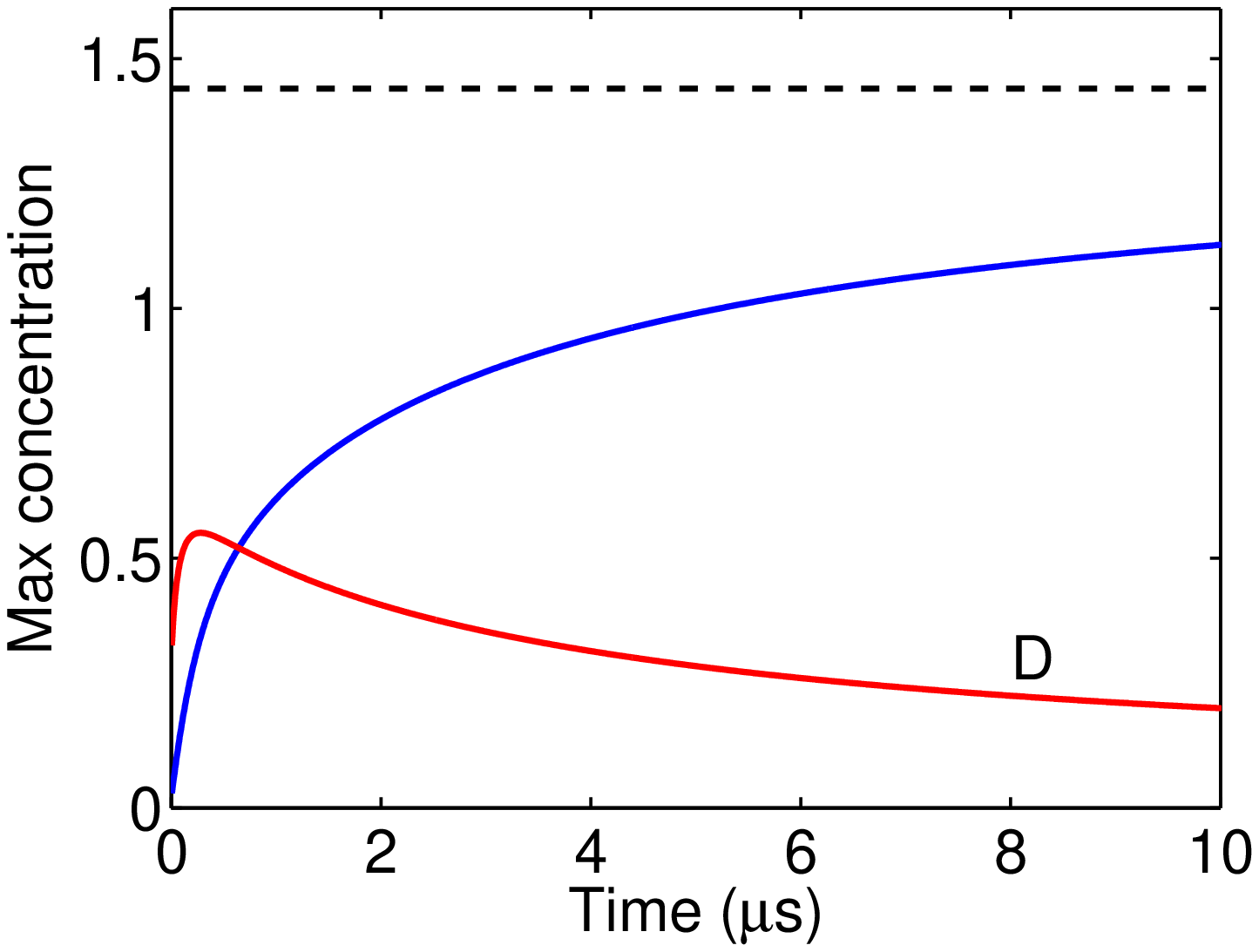}
\includegraphics[width=1.6in]{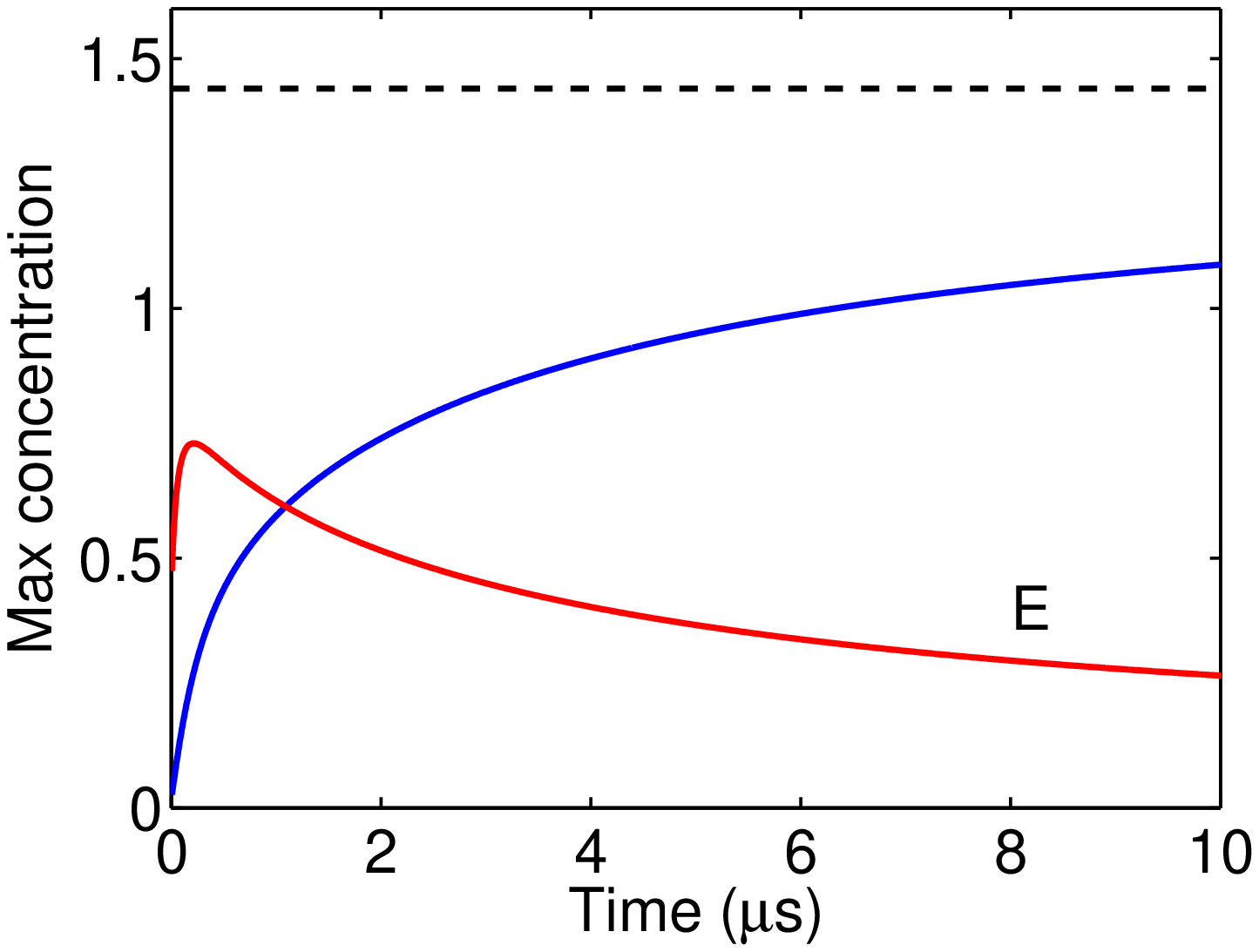}
\includegraphics[width=1.6in]{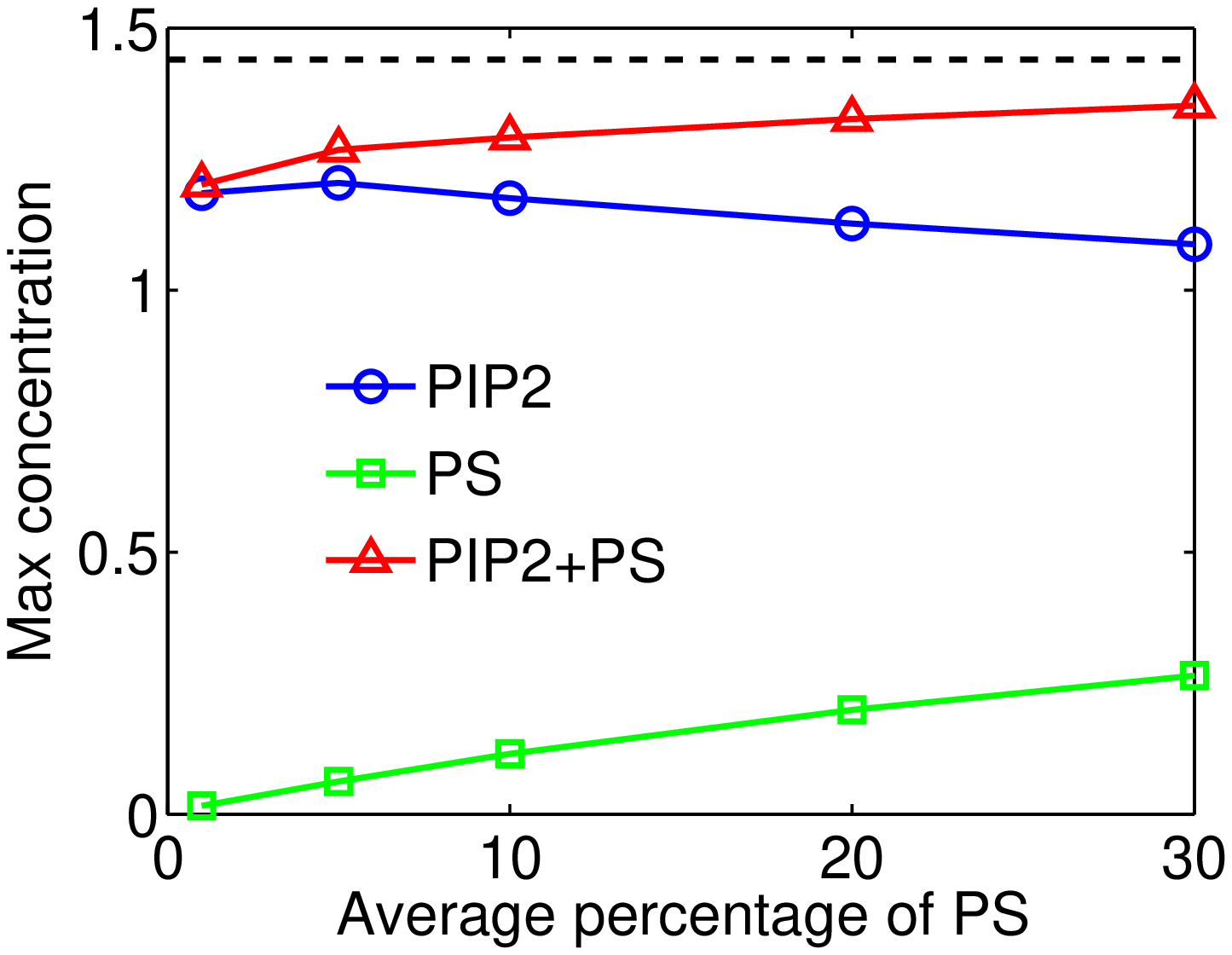}
\end{center}
\caption{
{\bf (A-E) Accumulation of \pip2 and PS near the MARCKS peptide and (F) the dependence of the maximum concentrations
of \pip2 and PS on the initial average concentration of PS}. The initial average concentrations of PS
in (A-E) are $1$\%, $5$\%, $10$\%, $20$\%, $30$\% respectively, with blue lines denoting PIP2 and red lines denoting
PS. The scale on the vertical axis is amplified 100-fold.
}
\label{fig:dionComp}
\end{figure}

When there are multiple species of negatively charged lipids on the membrane their concentrations
near the MARCKS peptide can be completely different. Boltzmann relation predicts that multivalent
lipids such as \pip2 have an affinity that is 1000 times stronger than the monovalent lipids
such as phosphatidylserine (PS) when binding to the MARCKS peptide. Although the real distribution
of charged lipids on a membrane does not follow the Boltzmann relation due to their large size
and correlations, electrostatic free energy of the sequestration showed that \pip2 is favored
by the peptide compared to PS \cite{GolebiewskaU2006a}. To quantify this competition we simulate electrodiffusion of
\pip2 and PS with different fractions. The average percentage of \pip2 is fixed at 1\% and the
fraction of PS varies between 1\% and 30\%. Figure \ref{fig:dionComp} shows the weak dependence
of the aggregation of \pip2 on the fractions of PS that are under investigation. While the total
concentrations of \pip2 and PS increase to approach the saturation value with the increase of
average concentration of PS, the peak concentration of \pip2 is found decreasing. This
decrease is slight nevertheless. The number of sequestered \pip2, computed using
equation (\ref{eqn:model_0}), reaches it smallest value of 2.7 when the average percentage of PS is 30\%;
the corresponding number of sequestered PS lipids is less than 0.4. These results are consistent with
the experimental observations that monovalent acidic lipids are less likely be sequestered by membrane-bound 
basic peptides if there are multivalent acidic lipids on the membrane \cite{GolebiewskaU2006a}, 
and that the sequestration of \pip2 is more favorable when the mol percentage of monovalent acidic lipids in the 
membrane decreases \cite{WangJ2004a}.

The time dependent solutions of electrodiffusion provides quantitative information that can not be
attained through the investigation of electrostatic free energy only. An examination of Figures
\ref{fig:dionComp}(C-E) shows that there is an increase of monovalent PS lipids for the first
few microseconds, in contrast to the continuous growth of the concentration of multivalent \pip2.
The strength of positive electrostatic potential near the sequestration region is reduced due to the
accumulation of negatively charged lipids. It appears that the monovalent PS lipids are more
susceptible to this local change potential so their concentration starts to decrease after a
maximum value is reached. This temporary increase of monovalent PS lipids is negligibly small
when their average percentage is low, c.f. Figure \ref{fig:dionComp}(A).

\subsection{Dependence of \pip2 aggregation on the ion concentration}
\begin{figure}[!ht]
\begin{center}
\includegraphics[width=3in]{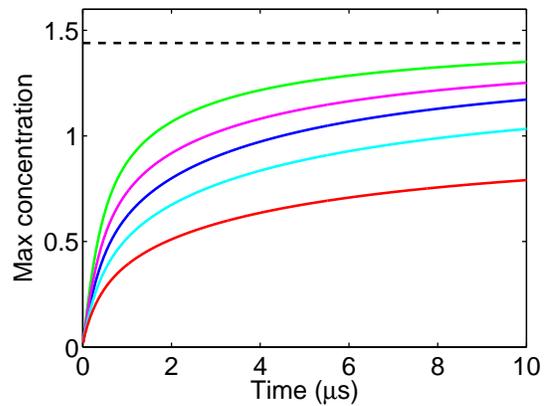}
\end{center}
\caption{
{\bf Salt-dependent lipid sequestration.} Maximum concentration of \pip2 decreases as salt concentration increases from
50mM (green), 100mM (magenta), 140mM (blue), 200mM (cyan) to 300mM (red). The number of sequestered \pip2 lipids are
4.02, 3.35, 2.89, 2.16 and 1.07, respectively.}
\label{fig:salt}
\end{figure}

Experiment observations suggest that the aggregation of charged lipids depends on the local 
electrostatic field \cite{Arbuzova1998a,WangJ2002a}, whose strength can change drastically with the ion concentration in the 
solution. This dependence is modeled in the current work through the coupling of the surface electrodiffusion equation with 
the Poisson equation (\ref{eqn:Poisson})
that admits the ion concentrations determined by the modified Nernst-Planck equations (\ref{eqn:NP_ion}). To validate the
model we simulate the \pip2 aggregation with KCl concentration ranging from 50mM to 300mM. Figure \ref{fig:salt} illustrates
that \pip2 lipids are less aggregated for large salt concentrations, as a result of the electrostatic screening and
the weakened tangent component $\nabla_s \phi$ of the electrostatic field. Our simulations also show that
other species of mobile ions such as Ca$^{2+}$, Mg$^{2+}$ will have similar screening effects. Note that
the Ca/CaM can also reverse the binding of MARCKS peptide to the membrane, so the current model
needs to be integrated with models for Ca/CaM association and Ca/CaM/MARCKS binding \cite{KimJ1994a} to
give a full description of the effects of Ca$^{2+}$ on \pip2 sequestration.

\section{Summary}
Lipid headgroups and membrane proteins diffuse laterally on the membrane surfaces. Local aggregation
of lipids and proteins is usually a precursor of membrane fusion, fission, and budding and therefore
is a critical step in many signaling pathways that are regulated by the membrane curvature. Lateral diffusion
of charged headgroups are mediated by
the nonspecific electrostatic interactions and hence can not be model by the conventional diffusion equation.
Here we establish a surface electrodiffusion model with finite-size correction
to describe the self-consistent lateral aggregation of charged lipid headgroups on curved membrane surfaces due to membrane/protein 
electrostatic interactions, without requiring the knowledge of locations of individual aggregated lipids. Our results
agree very well with the experimental observations and atomistic modeling of lipid sequestration. The ability to
efficiently and accurately identify the local concentrations of specified proteins or lipid headgroups will enable us
to integrate the self-consistent lateral translocation of lipids and proteins to the analysis of membrane trafficking
and curvature, and related signal pathways and diseases.


\begin{thebibliography}{52}%
\makeatletter
\providecommand \@ifxundefined [1]{%
 \@ifx{#1\undefined}
}%
\providecommand \@ifnum [1]{%
 \ifnum #1\expandafter \@firstoftwo
 \else \expandafter \@secondoftwo
 \fi
}%
\providecommand \@ifx [1]{%
 \ifx #1\expandafter \@firstoftwo
 \else \expandafter \@secondoftwo
 \fi
}%
\providecommand \natexlab [1]{#1}%
\providecommand \enquote  [1]{``#1''}%
\providecommand \bibnamefont  [1]{#1}%
\providecommand \bibfnamefont [1]{#1}%
\providecommand \citenamefont [1]{#1}%
\providecommand \href@noop [0]{\@secondoftwo}%
\providecommand \href [0]{\begingroup \@sanitize@url \@href}%
\providecommand \@href[1]{\@@startlink{#1}\@@href}%
\providecommand \@@href[1]{\endgroup#1\@@endlink}%
\providecommand \@sanitize@url [0]{\catcode `\\12\catcode `\$12\catcode
  `\&12\catcode `\#12\catcode `\^12\catcode `\_12\catcode `\%12\relax}%
\providecommand \@@startlink[1]{}%
\providecommand \@@endlink[0]{}%
\providecommand \url  [0]{\begingroup\@sanitize@url \@url }%
\providecommand \@url [1]{\endgroup\@href {#1}{\urlprefix }}%
\providecommand \urlprefix  [0]{URL }%
\providecommand \Eprint [0]{\href }%
\@ifxundefined \urlstyle {%
  \providecommand \doi  [0]{\begingroup \@sanitize@url \@doi}%
  \providecommand \@doi [1]{\endgroup \@@startlink {\doibase
  #1}doi:\discretionary {}{}{}#1\@@endlink }%
}{%
  \providecommand \doi  [0]{doi:\discretionary{}{}{}\begingroup
  \urlstyle{rm}\Url }%
}%
\providecommand \doibase [0]{http://dx.doi.org/}%
\providecommand \Doi [0]{\begingroup \@sanitize@url \@Doi }%
\providecommand \@Doi  [1]{\endgroup\@@startlink{\doibase#1}\@@Doi}%
\providecommand \@@Doi [1]{#1\@@endlink}%
\providecommand \selectlanguage [0]{\@gobble}%
\providecommand \bibinfo  [0]{\@secondoftwo}%
\providecommand \bibfield  [0]{\@secondoftwo}%
\providecommand \translation [1]{[#1]}%
\providecommand \BibitemOpen [0]{}%
\providecommand \bibitemStop [0]{}%
\providecommand \bibitemNoStop [0]{.\EOS\space}%
\providecommand \EOS [0]{\spacefactor3000\relax}%
\providecommand \BibitemShut  [1]{\csname bibitem#1\endcsname}%
\bibitem [{\citenamefont {McLaughlin}\ \emph {et~al.}(2002)\citenamefont
  {McLaughlin}, \citenamefont {Wang}, \citenamefont {Gambhir},\ and\
  \citenamefont {Murray}}]{McLaughlinS2002a}%
  \BibitemOpen
  \bibfield  {author} {\bibinfo {author} {\bibfnamefont {S.}~\bibnamefont
  {McLaughlin}}, \bibinfo {author} {\bibfnamefont {J.}~\bibnamefont {Wang}},
  \bibinfo {author} {\bibfnamefont {A.}~\bibnamefont {Gambhir}}, \ and\
  \bibinfo {author} {\bibfnamefont {D.}~\bibnamefont {Murray}},\ }\Doi
  {10.1146/annurev.biophys.31.082901.134259} {\bibfield  {journal} {\bibinfo
  {journal} {Annu.\ Rev.\ Biophys.\ Biomol.\ Struct.},\ }\textbf {\bibinfo
  {volume} {31}},\ \bibinfo {pages} {151} (\bibinfo {year} {2002})}\BibitemShut
  {NoStop}%
\bibitem [{\citenamefont {Suh}\ and\ \citenamefont {Hille}(2005)}]{SuhB2005a}%
  \BibitemOpen
  \bibfield  {author} {\bibinfo {author} {\bibfnamefont {B.-C.}\ \bibnamefont
  {Suh}}\ and\ \bibinfo {author} {\bibfnamefont {B.}~\bibnamefont {Hille}},\
  }\Doi {DOI: 10.1016/j.conb.2005.05.005} {\bibfield  {journal} {\bibinfo
  {journal} {Curr.\ Opin.\ Neurobiol.},\ }\textbf {\bibinfo {volume} {15}},\
  \bibinfo {pages} {370 } (\bibinfo {year} {2005})}\BibitemShut {NoStop}%
\bibitem [{\citenamefont {H\"{o}ning}\ \emph {et~al.}(2005)\citenamefont
  {H\"{o}ning}, \citenamefont {Ricotta}, \citenamefont {Krauss}, \citenamefont
  {Späte}, \citenamefont {Spolaore}, \citenamefont {Motley}, \citenamefont
  {Robinson}, \citenamefont {Robinson}, \citenamefont {Haucke},\ and\
  \citenamefont {Owen}}]{HoningS2005a}%
  \BibitemOpen
  \bibfield  {author} {\bibinfo {author} {\bibfnamefont {S.}~\bibnamefont
  {H\"{o}ning}}, \bibinfo {author} {\bibfnamefont {D.}~\bibnamefont {Ricotta}},
  \bibinfo {author} {\bibfnamefont {M.}~\bibnamefont {Krauss}}, \bibinfo
  {author} {\bibfnamefont {K.}~\bibnamefont {Späte}}, \bibinfo {author}
  {\bibfnamefont {B.}~\bibnamefont {Spolaore}}, \bibinfo {author}
  {\bibfnamefont {A.}~\bibnamefont {Motley}}, \bibinfo {author} {\bibfnamefont
  {M.}~\bibnamefont {Robinson}}, \bibinfo {author} {\bibfnamefont
  {C.}~\bibnamefont {Robinson}}, \bibinfo {author} {\bibfnamefont
  {V.}~\bibnamefont {Haucke}}, \ and\ \bibinfo {author} {\bibfnamefont {D.~J.}\
  \bibnamefont {Owen}},\ }\href@noop {} {\bibfield  {journal} {\bibinfo
  {journal} {Mol.\ Cell},\ }\textbf {\bibinfo {volume} {18}},\ \bibinfo {pages}
  {519} (\bibinfo {year} {2005})}\BibitemShut {NoStop}%
\bibitem [{\citenamefont {Janmey}\ and\ \citenamefont
  {Lindberg}(2004)}]{JanmeyP2004a}%
  \BibitemOpen
  \bibfield  {author} {\bibinfo {author} {\bibfnamefont {P.~A.}\ \bibnamefont
  {Janmey}}\ and\ \bibinfo {author} {\bibfnamefont {U.}~\bibnamefont
  {Lindberg}},\ }\Doi {doi:10.1038/nrm1434} {\bibfield  {journal} {\bibinfo
  {journal} {Nat.\ Rev.\ Mol.\ Cell Biol.},\ }\textbf {\bibinfo {volume} {5}},\
  \bibinfo {pages} {658} (\bibinfo {year} {2004})}\BibitemShut {NoStop}%
\bibitem [{\citenamefont {Blood}\ and\ \citenamefont
  {Voth}(2006)}]{BloodP2006}%
  \BibitemOpen
  \bibfield  {author} {\bibinfo {author} {\bibfnamefont {P.~D.}\ \bibnamefont
  {Blood}}\ and\ \bibinfo {author} {\bibfnamefont {G.~A.}\ \bibnamefont
  {Voth}},\ }\href@noop {} {\bibfield  {journal} {\bibinfo  {journal} {Proc.
  Natl. Acad. Sci. U.S.A.},\ }\textbf {\bibinfo {volume} {103}},\ \bibinfo
  {pages} {15068} (\bibinfo {year} {2006})}\BibitemShut {NoStop}%
\bibitem [{\citenamefont {McMahon}\ and\ \citenamefont
  {Gallop}(2005)}]{McMahonH2005}%
  \BibitemOpen
  \bibfield  {author} {\bibinfo {author} {\bibfnamefont {H.~T.}\ \bibnamefont
  {McMahon}}\ and\ \bibinfo {author} {\bibfnamefont {J.~L.}\ \bibnamefont
  {Gallop}},\ }\href@noop {} {\bibfield  {journal} {\bibinfo  {journal}
  {Nature},\ }\textbf {\bibinfo {volume} {438}},\ \bibinfo {pages} {590}
  (\bibinfo {year} {2005})}\BibitemShut {NoStop}%
\bibitem [{\citenamefont {Hinderliter}\ \emph {et~al.}(2001)\citenamefont
  {Hinderliter}, \citenamefont {Almeida}, \citenamefont {Creutz},\ and\
  \citenamefont {Biltonen}}]{HinderliterA2001a}%
  \BibitemOpen
  \bibfield  {author} {\bibinfo {author} {\bibfnamefont {A.}~\bibnamefont
  {Hinderliter}}, \bibinfo {author} {\bibfnamefont {P.~F.~F.}\ \bibnamefont
  {Almeida}}, \bibinfo {author} {\bibfnamefont {C.~E.}\ \bibnamefont {Creutz}},
  \ and\ \bibinfo {author} {\bibfnamefont {R.~L.}\ \bibnamefont {Biltonen}},\
  }\Doi {10.1021/bi0024299} {\bibfield  {journal} {\bibinfo  {journal}
  {Biochemistry},\ }\textbf {\bibinfo {volume} {40}},\ \bibinfo {pages} {4181}
  (\bibinfo {year} {2001})},\ \bibinfo {note} {pMID: 11300799}\BibitemShut
  {NoStop}%
\bibitem [{\citenamefont {McLaughlin}\ and\ \citenamefont
  {Murray}(2005)}]{MclaughlinS2005a}%
  \BibitemOpen
  \bibfield  {author} {\bibinfo {author} {\bibfnamefont {S.}~\bibnamefont
  {McLaughlin}}\ and\ \bibinfo {author} {\bibfnamefont {D.}~\bibnamefont
  {Murray}},\ }\Doi {10.1038/nature04398} {\bibfield  {journal} {\bibinfo
  {journal} {Nature},\ }\textbf {\bibinfo {volume} {438}},\ \bibinfo {pages}
  {605} (\bibinfo {year} {2005})}\BibitemShut {NoStop}%
\bibitem [{\citenamefont {Petrov}(2006)}]{PetrovA2006a}%
  \BibitemOpen
  \bibfield  {author} {\bibinfo {author} {\bibfnamefont {A.~G.}\ \bibnamefont
  {Petrov}},\ }\href@noop {} {\bibfield  {journal} {\bibinfo  {journal} {Curr.
  Top. Membr},\ }\textbf {\bibinfo {volume} {58}},\ \bibinfo {pages} {121}
  (\bibinfo {year} {2006})}\BibitemShut {NoStop}%
\bibitem [{\citenamefont {Filippov}\ \emph {et~al.}(2009)\citenamefont
  {Filippov}, \citenamefont {Orädd},\ and\ \citenamefont
  {Lindblom}}]{FilippovA2009a}%
  \BibitemOpen
  \bibfield  {author} {\bibinfo {author} {\bibfnamefont {A.}~\bibnamefont
  {Filippov}}, \bibinfo {author} {\bibfnamefont {G.}~\bibnamefont {Orädd}}, \
  and\ \bibinfo {author} {\bibfnamefont {G.}~\bibnamefont {Lindblom}},\ }\Doi
  {DOI: 10.1016/j.chemphyslip.2009.03.007} {\bibfield  {journal} {\bibinfo
  {journal} {Chemistry and Physics of Lipids},\ }\textbf {\bibinfo {volume}
  {159}},\ \bibinfo {pages} {81 } (\bibinfo {year} {2009})}\BibitemShut
  {NoStop}%
\bibitem [{\citenamefont {Fahey}\ \emph {et~al.}(1977)\citenamefont {Fahey},
  \citenamefont {Koppel}, \citenamefont {Barak}, \citenamefont {Wolf},
  \citenamefont {Elson},\ and\ \citenamefont {Webb}}]{FaheyP1977a}%
  \BibitemOpen
  \bibfield  {author} {\bibinfo {author} {\bibfnamefont {P.~F.}\ \bibnamefont
  {Fahey}}, \bibinfo {author} {\bibfnamefont {D.~E.}\ \bibnamefont {Koppel}},
  \bibinfo {author} {\bibfnamefont {L.~S.}\ \bibnamefont {Barak}}, \bibinfo
  {author} {\bibfnamefont {D.~E.}\ \bibnamefont {Wolf}}, \bibinfo {author}
  {\bibfnamefont {E.~L.}\ \bibnamefont {Elson}}, \ and\ \bibinfo {author}
  {\bibfnamefont {W.~W.}\ \bibnamefont {Webb}},\ }\Doi {DOI:
  10.1126/science.831279} {\bibfield  {journal} {\bibinfo  {journal}
  {Science},\ }\textbf {\bibinfo {volume} {195}},\ \bibinfo {pages} {305}
  (\bibinfo {year} {1977})}\BibitemShut {NoStop}%
\bibitem [{\citenamefont {Tanaka}\ \emph {et~al.}(1999)\citenamefont {Tanaka},
  \citenamefont {Manning}, \citenamefont {Lau},\ and\ \citenamefont
  {Yu}}]{TanakaK1999a}%
  \BibitemOpen
  \bibfield  {author} {\bibinfo {author} {\bibfnamefont {K.}~\bibnamefont
  {Tanaka}}, \bibinfo {author} {\bibfnamefont {P.~A.}\ \bibnamefont {Manning}},
  \bibinfo {author} {\bibfnamefont {V.~K.}\ \bibnamefont {Lau}}, \ and\
  \bibinfo {author} {\bibfnamefont {H.}~\bibnamefont {Yu}},\ }\href@noop {}
  {\bibfield  {journal} {\bibinfo  {journal} {Langmuir},\ }\textbf {\bibinfo
  {volume} {15}},\ \bibinfo {pages} {600} (\bibinfo {year} {1999})}\BibitemShut
  {NoStop}%
\bibitem [{\citenamefont {Ayton}\ and\ \citenamefont
  {Voth}(2004)}]{AytonG2004a}%
  \BibitemOpen
  \bibfield  {author} {\bibinfo {author} {\bibfnamefont {G.~S.}\ \bibnamefont
  {Ayton}}\ and\ \bibinfo {author} {\bibfnamefont {G.~A.}\ \bibnamefont
  {Voth}},\ }\href@noop {} {\bibfield  {journal} {\bibinfo  {journal} {Biophys.
  J.},\ }\textbf {\bibinfo {volume} {87}},\ \bibinfo {pages} {3299} (\bibinfo
  {year} {2004})}\BibitemShut {NoStop}%
\bibitem [{\citenamefont {Kuo}\ and\ \citenamefont {Wade}(1979)}]{KuoA1979a}%
  \BibitemOpen
  \bibfield  {author} {\bibinfo {author} {\bibfnamefont {A.-L.}\ \bibnamefont
  {Kuo}}\ and\ \bibinfo {author} {\bibfnamefont {C.~G.}\ \bibnamefont {Wade}},\
  }\Doi {DOI: 10.1021/bi00578a026} {\bibfield  {journal} {\bibinfo  {journal}
  {Biochemistry},\ }\textbf {\bibinfo {volume} {18}},\ \bibinfo {pages} {2300}
  (\bibinfo {year} {1979})}\BibitemShut {NoStop}%
\bibitem [{\citenamefont {Ellena}\ \emph {et~al.}(1993)\citenamefont {Ellena},
  \citenamefont {Lepore},\ and\ \citenamefont {Cafiso}}]{EllensJ1993a}%
  \BibitemOpen
  \bibfield  {author} {\bibinfo {author} {\bibfnamefont {J.~F.}\ \bibnamefont
  {Ellena}}, \bibinfo {author} {\bibfnamefont {L.~S.}\ \bibnamefont {Lepore}},
  \ and\ \bibinfo {author} {\bibfnamefont {D.~S.}\ \bibnamefont {Cafiso}},\
  }\Doi {doi:10.1021/j100114a021} {\bibfield  {journal} {\bibinfo  {journal}
  {J.\ Phys.\ Chem.},\ }\textbf {\bibinfo {volume} {97}},\ \bibinfo {pages}
  {2952} (\bibinfo {year} {1993})}\BibitemShut {NoStop}%
\bibitem [{\citenamefont {Cicuta}\ \emph {et~al.}(2007)\citenamefont {Cicuta},
  \citenamefont {Keller},\ and\ \citenamefont {Veatch}}]{CicutaP2007a}%
  \BibitemOpen
  \bibfield  {author} {\bibinfo {author} {\bibfnamefont {P.}~\bibnamefont
  {Cicuta}}, \bibinfo {author} {\bibfnamefont {S.~L.}\ \bibnamefont {Keller}},
  \ and\ \bibinfo {author} {\bibfnamefont {S.~L.}\ \bibnamefont {Veatch}},\
  }\Doi {10.1021/jp0702088} {\bibfield  {journal} {\bibinfo  {journal} {J.\
  Phys.\ Chem.\ B},\ }\textbf {\bibinfo {volume} {111}},\ \bibinfo {pages}
  {3328} (\bibinfo {year} {2007})}\BibitemShut {NoStop}%
\bibitem [{\citenamefont {Chen}\ \emph {et~al.}(2006)\citenamefont {Chen},
  \citenamefont {Lagerholm}, \citenamefont {Yang},\ and\ \citenamefont
  {Jacobson}}]{ChenY2006a}%
  \BibitemOpen
  \bibfield  {author} {\bibinfo {author} {\bibfnamefont {Y.}~\bibnamefont
  {Chen}}, \bibinfo {author} {\bibfnamefont {B.~C.}\ \bibnamefont {Lagerholm}},
  \bibinfo {author} {\bibfnamefont {B.}~\bibnamefont {Yang}}, \ and\ \bibinfo
  {author} {\bibfnamefont {K.}~\bibnamefont {Jacobson}},\ }\Doi {DOI:
  10.1016/j.ymeth.2006.05.008} {\bibfield  {journal} {\bibinfo  {journal}
  {Methods},\ }\textbf {\bibinfo {volume} {39}},\ \bibinfo {pages} {147 }
  (\bibinfo {year} {2006})},\ ISSN \bibinfo {issn} {1046-2023},\ \bibinfo
  {note} {analytical methods in the science of lipidomics, membrane
  organization and Protein-Lipid interactions}\BibitemShut {NoStop}%
\bibitem [{\citenamefont {Tocanne}\ \emph {et~al.}(1989)\citenamefont
  {Tocanne}, \citenamefont {Dupou-Cézanne}, \citenamefont {Lopez},\ and\
  \citenamefont {Tournier}}]{TocanneJ1989a}%
  \BibitemOpen
  \bibfield  {author} {\bibinfo {author} {\bibfnamefont {J.~F.}\ \bibnamefont
  {Tocanne}}, \bibinfo {author} {\bibfnamefont {L.}~\bibnamefont
  {Dupou-Cézanne}}, \bibinfo {author} {\bibfnamefont {A.}~\bibnamefont
  {Lopez}}, \ and\ \bibinfo {author} {\bibfnamefont {J.~F.}\ \bibnamefont
  {Tournier}},\ }\Doi {doi:10.1016/0014-5793(89)81774-0} {\bibfield  {journal}
  {\bibinfo  {journal} {FEBS Lett.},\ }\textbf {\bibinfo {volume} {257}},\
  \bibinfo {pages} {10} (\bibinfo {year} {1989})}\BibitemShut {NoStop}%
\bibitem [{\citenamefont {Vaz}\ and\ \citenamefont
  {Almeida}(1991)}]{VazW1991a}%
  \BibitemOpen
  \bibfield  {author} {\bibinfo {author} {\bibfnamefont {W.~L.~C.}\
  \bibnamefont {Vaz}}\ and\ \bibinfo {author} {\bibfnamefont {P.~F.~F.}\
  \bibnamefont {Almeida}},\ }\Doi {10.1016/S0006-3495(91)82190-7} {\bibfield
  {journal} {\bibinfo  {journal} {Biophys. J.},\ }\textbf {\bibinfo {volume}
  {60}},\ \bibinfo {pages} {1553} (\bibinfo {year} {1991})}\BibitemShut
  {NoStop}%
\bibitem [{\citenamefont {Vattulainen}\ and\ \citenamefont
  {Mouritsen}(2005)}]{VattulainenI1005a}%
  \BibitemOpen
  \bibfield  {author} {\bibinfo {author} {\bibfnamefont {I.}~\bibnamefont
  {Vattulainen}}\ and\ \bibinfo {author} {\bibfnamefont {O.~G.}\ \bibnamefont
  {Mouritsen}},\ }in\ \href@noop {} {\emph {\bibinfo {booktitle} {Diffusion in
  Condensed Matter: Methods, Materials, Models}}}\ (\bibinfo  {publisher}
  {Springer-Verlag},\ \bibinfo {year} {2005})\ pp.\ \bibinfo {pages}
  {471--509}\BibitemShut {NoStop}%
\bibitem [{\citenamefont {Cohen}\ and\ \citenamefont
  {Turnbull}(1959)}]{CohenM1959a}%
  \BibitemOpen
  \bibfield  {author} {\bibinfo {author} {\bibfnamefont {M.}~\bibnamefont
  {Cohen}}\ and\ \bibinfo {author} {\bibfnamefont {D.}~\bibnamefont
  {Turnbull}},\ }\href@noop {} {\bibfield  {journal} {\bibinfo  {journal} {J.
  Chem. Phys.},\ }\textbf {\bibinfo {volume} {31}},\ \bibinfo {pages} {1164}
  (\bibinfo {year} {1959})}\BibitemShut {NoStop}%
\bibitem [{\citenamefont {Turnbull}\ and\ \citenamefont
  {Cohen}(1961)}]{TurnbullD1961a}%
  \BibitemOpen
  \bibfield  {author} {\bibinfo {author} {\bibfnamefont {D.}~\bibnamefont
  {Turnbull}}\ and\ \bibinfo {author} {\bibfnamefont {M.}~\bibnamefont
  {Cohen}},\ }\href@noop {} {\bibfield  {journal} {\bibinfo  {journal} {J.
  Chem. Phys.},\ }\textbf {\bibinfo {volume} {34}},\ \bibinfo {pages} {120}
  (\bibinfo {year} {1961})}\BibitemShut {NoStop}%
\bibitem [{\citenamefont {MacCarthy}\ and\ \citenamefont
  {Kozak}(1982)}]{MaccarthyJ1982a}%
  \BibitemOpen
  \bibfield  {author} {\bibinfo {author} {\bibfnamefont {J.~E.}\ \bibnamefont
  {MacCarthy}}\ and\ \bibinfo {author} {\bibfnamefont {J.~J.}\ \bibnamefont
  {Kozak}},\ }\href@noop {} {\bibfield  {journal} {\bibinfo  {journal} {J.
  Chem. Phys.},\ }\textbf {\bibinfo {volume} {77}},\ \bibinfo {pages} {2214}
  (\bibinfo {year} {1982})}\BibitemShut {NoStop}%
\bibitem [{\citenamefont {Saxton}(1982)}]{SaxtonM1982a}%
  \BibitemOpen
  \bibfield  {author} {\bibinfo {author} {\bibfnamefont {M.~J.}\ \bibnamefont
  {Saxton}},\ }\href@noop {} {\bibfield  {journal} {\bibinfo  {journal}
  {Biophys. J.},\ }\textbf {\bibinfo {volume} {39}},\ \bibinfo {pages} {165}
  (\bibinfo {year} {1982})}\BibitemShut {NoStop}%
\bibitem [{\citenamefont {Lebowitz}\ \emph {et~al.}(1965)\citenamefont
  {Lebowitz}, \citenamefont {Helfand},\ and\ \citenamefont
  {Praestgaard}}]{LebowitzJ1965a}%
  \BibitemOpen
  \bibfield  {author} {\bibinfo {author} {\bibfnamefont {J.~L.}\ \bibnamefont
  {Lebowitz}}, \bibinfo {author} {\bibfnamefont {E.}~\bibnamefont {Helfand}}, \
  and\ \bibinfo {author} {\bibfnamefont {E.}~\bibnamefont {Praestgaard}},\
  }\href@noop {} {\bibfield  {journal} {\bibinfo  {journal} {J. Chem. Phys.},\
  }\textbf {\bibinfo {volume} {43}},\ \bibinfo {pages} {774} (\bibinfo {year}
  {1965})}\BibitemShut {NoStop}%
\bibitem [{\citenamefont {O'leary}(1987)}]{OlearyJ1984a}%
  \BibitemOpen
  \bibfield  {author} {\bibinfo {author} {\bibfnamefont {T.~J.}\ \bibnamefont
  {O'leary}},\ }\Doi {DOI:10.1073/pnas.84.2.429} {\bibfield  {journal}
  {\bibinfo  {journal} {Proc. Natl. Acad. Sci. U.S.A.},\ }\textbf {\bibinfo
  {volume} {84}},\ \bibinfo {pages} {429} (\bibinfo {year} {1987})}\BibitemShut
  {NoStop}%
\bibitem [{\citenamefont {Falck}\ \emph {et~al.}(2008)\citenamefont {Falck},
  \citenamefont {Róg}, \citenamefont {Karttunen},\ and\ \citenamefont
  {Vattulainen}}]{FalckE2008a}%
  \BibitemOpen
  \bibfield  {author} {\bibinfo {author} {\bibfnamefont {E.}~\bibnamefont
  {Falck}}, \bibinfo {author} {\bibfnamefont {T.}~\bibnamefont {Róg}},
  \bibinfo {author} {\bibfnamefont {M.}~\bibnamefont {Karttunen}}, \ and\
  \bibinfo {author} {\bibfnamefont {I.}~\bibnamefont {Vattulainen}},\ }\Doi
  {http://dx.doi.org/10.1021/ja7103558} {\bibfield  {journal} {\bibinfo
  {journal} {J. Am. Chem. Soc.},\ }\textbf {\bibinfo {volume} {130}},\ \bibinfo
  {pages} {44} (\bibinfo {year} {2008})}\BibitemShut {NoStop}%
\bibitem [{\citenamefont {Amatore}\ \emph
  {et~al.}(2009){\natexlab{a}}\citenamefont {Amatore}, \citenamefont
  {Oleinick}, \citenamefont {Klymenko},\ and\ \citenamefont
  {Svir}}]{AmatoreC2009a}%
  \BibitemOpen
  \bibfield  {author} {\bibinfo {author} {\bibfnamefont {C.}~\bibnamefont
  {Amatore}}, \bibinfo {author} {\bibfnamefont {A.~I.}\ \bibnamefont
  {Oleinick}}, \bibinfo {author} {\bibfnamefont {O.~V.}\ \bibnamefont
  {Klymenko}}, \ and\ \bibinfo {author} {\bibfnamefont {I.}~\bibnamefont
  {Svir}},\ }\href@noop {} {\bibfield  {journal} {\bibinfo  {journal}
  {Chemphyschem.},\ }\textbf {\bibinfo {volume} {10}},\ \bibinfo {pages} {1586}
  (\bibinfo {year} {2009}{\natexlab{a}})}\BibitemShut {NoStop}%
\bibitem [{\citenamefont {Amatore}\ \emph
  {et~al.}(2009){\natexlab{b}}\citenamefont {Amatore}, \citenamefont
  {Klymenko}, \citenamefont {Oleinick},\ and\ \citenamefont
  {Svir}}]{AmatoreC2009b}%
  \BibitemOpen
  \bibfield  {author} {\bibinfo {author} {\bibfnamefont {C.}~\bibnamefont
  {Amatore}}, \bibinfo {author} {\bibfnamefont {O.~V.}\ \bibnamefont
  {Klymenko}}, \bibinfo {author} {\bibfnamefont {A.~I.}\ \bibnamefont
  {Oleinick}}, \ and\ \bibinfo {author} {\bibfnamefont {I.}~\bibnamefont
  {Svir}},\ }\href@noop {} {\bibfield  {journal} {\bibinfo  {journal}
  {Chemphyschem.},\ }\textbf {\bibinfo {volume} {10}},\ \bibinfo {pages} {1593}
  (\bibinfo {year} {2009}{\natexlab{b}})}\BibitemShut {NoStop}%
\bibitem [{\citenamefont {Yoshigaki}(2007)}]{YoshigakiT2007a}%
  \BibitemOpen
  \bibfield  {author} {\bibinfo {author} {\bibfnamefont {T.}~\bibnamefont
  {Yoshigaki}},\ }\Doi {10.1103/PhysRevE.75.041901} {\bibfield  {journal}
  {\bibinfo  {journal} {Phys. Rev. E},\ }\textbf {\bibinfo {volume} {75}},\
  \bibinfo {pages} {041901} (\bibinfo {year} {2007})}\BibitemShut {NoStop}%
\bibitem [{\citenamefont {Leitenberger}\ \emph {et~al.}(2008)\citenamefont
  {Leitenberger}, \citenamefont {Reister-Gottfried},\ and\ \citenamefont
  {Seifert}}]{LeitenbergerS2008a}%
  \BibitemOpen
  \bibfield  {author} {\bibinfo {author} {\bibfnamefont {S.~M.}\ \bibnamefont
  {Leitenberger}}, \bibinfo {author} {\bibfnamefont {E.}~\bibnamefont
  {Reister-Gottfried}}, \ and\ \bibinfo {author} {\bibfnamefont
  {U.}~\bibnamefont {Seifert}},\ }\Doi {10.1021/la702319q} {\bibfield
  {journal} {\bibinfo  {journal} {Langmuir},\ }\textbf {\bibinfo {volume}
  {24}},\ \bibinfo {pages} {1254} (\bibinfo {year} {2008})}\BibitemShut
  {NoStop}%
\bibitem [{\citenamefont {G\'o\'z\'d\'z}(2008)}]{GozdaW2008a}%
  \BibitemOpen
  \bibfield  {author} {\bibinfo {author} {\bibfnamefont {W.~T.}\ \bibnamefont
  {G\'o\'z\'d\'z}},\ }\Doi {10.1021/la801767q} {\bibfield  {journal} {\bibinfo
  {journal} {Langmuir},\ }\textbf {\bibinfo {volume} {24}},\ \bibinfo {pages}
  {12458} (\bibinfo {year} {2008})}\BibitemShut {NoStop}%
\bibitem [{\citenamefont {Chevalier}\ and\ \citenamefont
  {Debbasch}(2007)}]{ChevalierC2007a}%
  \BibitemOpen
  \bibfield  {author} {\bibinfo {author} {\bibfnamefont {C.}~\bibnamefont
  {Chevalier}}\ and\ \bibinfo {author} {\bibfnamefont {F.}~\bibnamefont
  {Debbasch}},\ }\Doi {10.1209/0295-5075/77/20005} {\bibfield  {journal}
  {\bibinfo  {journal} {Europhys.\ Lett.},\ }\textbf {\bibinfo {volume} {77}},\
  \bibinfo {pages} {20005} (\bibinfo {year} {2007})}\BibitemShut {NoStop}%
\bibitem [{\citenamefont {Kilic}\ \emph {et~al.}(2007)\citenamefont {Kilic},
  \citenamefont {Bazant},\ and\ \citenamefont {Ajdari}}]{Kilic2007b}%
  \BibitemOpen
  \bibfield  {author} {\bibinfo {author} {\bibfnamefont {M.~S.}\ \bibnamefont
  {Kilic}}, \bibinfo {author} {\bibfnamefont {M.~Z.}\ \bibnamefont {Bazant}}, \
  and\ \bibinfo {author} {\bibfnamefont {A.}~\bibnamefont {Ajdari}},\
  }\href@noop {} {\bibfield  {journal} {\bibinfo  {journal} {Phys. Rev. E},\
  }\textbf {\bibinfo {volume} {75}},\ \bibinfo {eid} {021503} (\bibinfo {year}
  {2007})}\BibitemShut {NoStop}%
\bibitem [{\citenamefont {Lu}\ and\ \citenamefont {Zhou}(2011)}]{ZhouY2011a}%
  \BibitemOpen
  \bibfield  {author} {\bibinfo {author} {\bibfnamefont {B.}~\bibnamefont
  {Lu}}\ and\ \bibinfo {author} {\bibfnamefont {Y.~C.}\ \bibnamefont {Zhou}},\
  }\Doi {10.1016/j.bpj.2011.03.059} {\bibfield  {journal} {\bibinfo  {journal}
  {Biophys. J.},\ }\textbf {\bibinfo {volume} {100}},\ \bibinfo {pages} {2475}
  (\bibinfo {year} {2011})}\BibitemShut {NoStop}%
\bibitem [{\citenamefont {Kiselev}\ \emph
  {et~al.}(2011){\natexlab{a}}\citenamefont {Kiselev}, \citenamefont {Leda},
  \citenamefont {Lobanov}, \citenamefont {Marenduzzo},\ and\ \citenamefont
  {Goryachev}}]{KiselevV2011a}%
  \BibitemOpen
  \bibfield  {author} {\bibinfo {author} {\bibfnamefont {V.~Y.}\ \bibnamefont
  {Kiselev}}, \bibinfo {author} {\bibfnamefont {M.}~\bibnamefont {Leda}},
  \bibinfo {author} {\bibfnamefont {A.~I.}\ \bibnamefont {Lobanov}}, \bibinfo
  {author} {\bibfnamefont {D.}~\bibnamefont {Marenduzzo}}, \ and\ \bibinfo
  {author} {\bibfnamefont {A.~B.}\ \bibnamefont {Goryachev}},\ }\Doi
  {10.1063/1.3652958} {\bibfield  {journal} {\bibinfo  {journal} {J. Chem.
  Phys.},\ }\textbf {\bibinfo {volume} {135}},\ \bibinfo {pages} {155103}
  (\bibinfo {year} {2011}{\natexlab{a}})}\BibitemShut {NoStop}%
\bibitem [{\citenamefont {Kiselev}\ \emph
  {et~al.}(2011){\natexlab{b}}\citenamefont {Kiselev}, \citenamefont
  {Marenduzzo},\ and\ \citenamefont {Goryachev}}]{KiselevV2011b}%
  \BibitemOpen
  \bibfield  {author} {\bibinfo {author} {\bibfnamefont {V.~Y.}\ \bibnamefont
  {Kiselev}}, \bibinfo {author} {\bibfnamefont {D.}~\bibnamefont {Marenduzzo}},
  \ and\ \bibinfo {author} {\bibfnamefont {A.~B.}\ \bibnamefont {Goryachev}},\
  }\Doi {10.1016/j.bpj.2011.01.025} {\bibfield  {journal} {\bibinfo  {journal}
  {Biophys. J.},\ }\textbf {\bibinfo {volume} {100}},\ \bibinfo {pages} {1261}
  (\bibinfo {year} {2011}{\natexlab{b}})}\BibitemShut {NoStop}%
\bibitem [{\citenamefont {Khelashvill}\ \emph {et~al.}(2008)\citenamefont
  {Khelashvill}, \citenamefont {Weinstein},\ and\ \citenamefont
  {Harries}}]{KhelashviliG2008a}%
  \BibitemOpen
  \bibfield  {author} {\bibinfo {author} {\bibfnamefont {G.}~\bibnamefont
  {Khelashvill}}, \bibinfo {author} {\bibfnamefont {H.}~\bibnamefont
  {Weinstein}}, \ and\ \bibinfo {author} {\bibfnamefont {D.}~\bibnamefont
  {Harries}},\ }\Doi {10.1529/biophysj.107.120667} {\bibfield  {journal}
  {\bibinfo  {journal} {Biophys. J.},\ }\textbf {\bibinfo {volume} {94}},\
  \bibinfo {pages} {2580} (\bibinfo {year} {2008})}\BibitemShut {NoStop}%
\bibitem [{\citenamefont {Borukhov}\ \emph {et~al.}(1997)\citenamefont
  {Borukhov}, \citenamefont {Andelman},\ and\ \citenamefont
  {Orland}}]{Borukhov1997}%
  \BibitemOpen
  \bibfield  {author} {\bibinfo {author} {\bibfnamefont {I.}~\bibnamefont
  {Borukhov}}, \bibinfo {author} {\bibfnamefont {D.}~\bibnamefont {Andelman}},
  \ and\ \bibinfo {author} {\bibfnamefont {H.}~\bibnamefont {Orland}},\ }\Doi
  {10.1103/PhysRevLett.79.435} {\bibfield  {journal} {\bibinfo  {journal}
  {Phys.\ Rev.\ Lett.},\ }\textbf {\bibinfo {volume} {79}},\ \bibinfo {pages}
  {435} (\bibinfo {year} {1997})}\BibitemShut {NoStop}%
\bibitem [{\citenamefont {Burak}\ and\ \citenamefont
  {Andelman}(2001)}]{Burak01}%
  \BibitemOpen
  \bibfield  {author} {\bibinfo {author} {\bibfnamefont {Y.}~\bibnamefont
  {Burak}}\ and\ \bibinfo {author} {\bibfnamefont {D.}~\bibnamefont
  {Andelman}},\ }\Doi {10.1063/1.1331569} {\bibfield  {journal} {\bibinfo
  {journal} {J. Chem. Phys.},\ }\textbf {\bibinfo {volume} {114}},\ \bibinfo
  {pages} {3271} (\bibinfo {year} {2001})}\BibitemShut {NoStop}%
\bibitem [{\citenamefont {Hansen}\ and\ \citenamefont
  {McDonald}(2006)}]{theoryofsimpleliquids}%
  \BibitemOpen
  \bibfield  {author} {\bibinfo {author} {\bibfnamefont {J.-P.}\ \bibnamefont
  {Hansen}}\ and\ \bibinfo {author} {\bibfnamefont {I.~R.}\ \bibnamefont
  {McDonald}},\ }\href@noop {} {\emph {\bibinfo {title} {Theory of simple
  liquids}}}\ (\bibinfo  {publisher} {Academic Press},\ \bibinfo {year}
  {2006})\BibitemShut {NoStop}%
\bibitem [{\citenamefont {Stone}(1990)}]{StoneH1990a}%
  \BibitemOpen
  \bibfield  {author} {\bibinfo {author} {\bibfnamefont {H.~A.}\ \bibnamefont
  {Stone}},\ }\href@noop {} {\bibfield  {journal} {\bibinfo  {journal} {Phys.
  Fluids A},\ }\textbf {\bibinfo {volume} {2}},\ \bibinfo {pages} {111}
  (\bibinfo {year} {1990})}\BibitemShut {NoStop}%
\bibitem [{\citenamefont {Dziuk}(1988)}]{Dziukg1988a}%
  \BibitemOpen
  \bibfield  {author} {\bibinfo {author} {\bibfnamefont {G.}~\bibnamefont
  {Dziuk}},\ }in\ \href@noop {} {\emph {\bibinfo {booktitle} {Partial
  differential equations and calculus of variations}}}\ (\bibinfo  {publisher}
  {Springer, Berlin},\ \bibinfo {year} {1988})\BibitemShut {NoStop}%
\bibitem [{\citenamefont {Arbuzova}\ \emph {et~al.}(2002)\citenamefont
  {Arbuzova}, \citenamefont {Schmitz},\ and\ \citenamefont
  {Ver\`eres}}]{ArbuzovaA2002a}%
  \BibitemOpen
  \bibfield  {author} {\bibinfo {author} {\bibfnamefont {A.}~\bibnamefont
  {Arbuzova}}, \bibinfo {author} {\bibfnamefont {A.~A.~P.}\ \bibnamefont
  {Schmitz}}, \ and\ \bibinfo {author} {\bibfnamefont {G.}~\bibnamefont
  {Ver\`eres}},\ }\Doi {10.1042/0264-6021:3620001} {\bibfield  {journal}
  {\bibinfo  {journal} {Biochem. J.},\ }\textbf {\bibinfo {volume} {15}},\
  \bibinfo {pages} {1} (\bibinfo {year} {2002})}\BibitemShut {NoStop}%
\bibitem [{\citenamefont {Ratto}\ and\ \citenamefont
  {Longo}(2002)}]{RattoT2002a}%
  \BibitemOpen
  \bibfield  {author} {\bibinfo {author} {\bibfnamefont {T.~V.}\ \bibnamefont
  {Ratto}}\ and\ \bibinfo {author} {\bibfnamefont {M.~L.}\ \bibnamefont
  {Longo}},\ }\href@noop {} {\bibfield  {journal} {\bibinfo  {journal}
  {Biophys. J.},\ }\textbf {\bibinfo {volume} {83}},\ \bibinfo {pages} {3380}
  (\bibinfo {year} {2002})}\BibitemShut {NoStop}%
\bibitem [{\citenamefont {Shorten}\ and\ \citenamefont
  {Sneyd}(2009)}]{ShortenP2009a}%
  \BibitemOpen
  \bibfield  {author} {\bibinfo {author} {\bibfnamefont {P.}~\bibnamefont
  {Shorten}}\ and\ \bibinfo {author} {\bibfnamefont {J.}~\bibnamefont
  {Sneyd}},\ }\Doi {10.1016/j.bpj.2009.02.060} {\bibfield  {journal} {\bibinfo
  {journal} {Biophys. J.},\ }\textbf {\bibinfo {volume} {96}},\ \bibinfo
  {pages} {4764} (\bibinfo {year} {2009})}\BibitemShut {NoStop}%
\bibitem [{\citenamefont {Flenner}\ \emph {et~al.}(2009)\citenamefont
  {Flenner}, \citenamefont {Das}, \citenamefont {Rheinst\"adter},\ and\
  \citenamefont {Kosztin}}]{FlennerE2009a}%
  \BibitemOpen
  \bibfield  {author} {\bibinfo {author} {\bibfnamefont {E.}~\bibnamefont
  {Flenner}}, \bibinfo {author} {\bibfnamefont {J.}~\bibnamefont {Das}},
  \bibinfo {author} {\bibfnamefont {M.~C.}\ \bibnamefont {Rheinst\"adter}}, \
  and\ \bibinfo {author} {\bibfnamefont {I.}~\bibnamefont {Kosztin}},\ }\Doi
  {10.1103/PhysRevE.79.011907} {\bibfield  {journal} {\bibinfo  {journal}
  {Phys.\ Rev.\ E},\ }\textbf {\bibinfo {volume} {79}},\ \bibinfo {pages}
  {011907} (\bibinfo {year} {2009})}\BibitemShut {NoStop}%
\bibitem [{\citenamefont {Zhou}\ \emph {et~al.}(2010)\citenamefont {Zhou},
  \citenamefont {Lu},\ and\ \citenamefont {Gorfe}}]{ZhouY2010b}%
  \BibitemOpen
  \bibfield  {author} {\bibinfo {author} {\bibfnamefont {Y.~C.}\ \bibnamefont
  {Zhou}}, \bibinfo {author} {\bibfnamefont {B.}~\bibnamefont {Lu}}, \ and\
  \bibinfo {author} {\bibfnamefont {A.~A.}\ \bibnamefont {Gorfe}},\ }\Doi
  {10.1103/PhysRevE.82.041923} {\bibfield  {journal} {\bibinfo  {journal}
  {Phys. Rev. E},\ }\textbf {\bibinfo {volume} {82}},\ \bibinfo {pages}
  {041923} (\bibinfo {year} {2010})}\BibitemShut {NoStop}%
\bibitem [{\citenamefont {Wang}\ \emph {et~al.}(2002)\citenamefont {Wang},
  \citenamefont {Gambhir}, \citenamefont {lyne}, \citenamefont {Murray},
  \citenamefont {Golebiewska},\ and\ \citenamefont {McLaughlin}}]{WangJ2002a}%
  \BibitemOpen
  \bibfield  {author} {\bibinfo {author} {\bibfnamefont {J.}~\bibnamefont
  {Wang}}, \bibinfo {author} {\bibfnamefont {A.}~\bibnamefont {Gambhir}},
  \bibinfo {author} {\bibfnamefont {G.~H.-M.}\ \bibnamefont {lyne}}, \bibinfo
  {author} {\bibfnamefont {D.}~\bibnamefont {Murray}}, \bibinfo {author}
  {\bibfnamefont {U.}~\bibnamefont {Golebiewska}}, \ and\ \bibinfo {author}
  {\bibfnamefont {S.}~\bibnamefont {McLaughlin}},\ }\Doi
  {10.1074/jbc.M203954200} {\bibfield  {journal} {\bibinfo  {journal} {J.\
  Biol.\ Chem.},\ }\textbf {\bibinfo {volume} {277}},\ \bibinfo {pages} {34401}
  (\bibinfo {year} {2002})}\BibitemShut {NoStop}%
\bibitem [{\citenamefont {Wang}\ \emph {et~al.}(2004)\citenamefont {Wang},
  \citenamefont {Gambhir}, \citenamefont {McLaughlin},\ and\ \citenamefont
  {Murray}}]{WangJ2004a}%
  \BibitemOpen
  \bibfield  {author} {\bibinfo {author} {\bibfnamefont {J.}~\bibnamefont
  {Wang}}, \bibinfo {author} {\bibfnamefont {A.}~\bibnamefont {Gambhir}},
  \bibinfo {author} {\bibfnamefont {S.}~\bibnamefont {McLaughlin}}, \ and\
  \bibinfo {author} {\bibfnamefont {D.}~\bibnamefont {Murray}},\ }\Doi
  {10.1016/S0006-3495(04)74260-5} {\bibfield  {journal} {\bibinfo  {journal}
  {Biophys. J.},\ }\textbf {\bibinfo {volume} {86}},\ \bibinfo {pages} {1969}
  (\bibinfo {year} {2004})}\BibitemShut {NoStop}%
\bibitem [{\citenamefont {Golebiewska}\ \emph {et~al.}(2006)\citenamefont
  {Golebiewska}, \citenamefont {Gambhir}, \citenamefont {Hangyás-Mihályné},
  \citenamefont {Zaitseva}, \citenamefont {Rädler},\ and\ \citenamefont
  {McLaughlin}}]{GolebiewskaU2006a}%
  \BibitemOpen
  \bibfield  {author} {\bibinfo {author} {\bibfnamefont {U.}~\bibnamefont
  {Golebiewska}}, \bibinfo {author} {\bibfnamefont {A.}~\bibnamefont
  {Gambhir}}, \bibinfo {author} {\bibfnamefont {G.}~\bibnamefont
  {Hangyás-Mihályné}}, \bibinfo {author} {\bibfnamefont {I.}~\bibnamefont
  {Zaitseva}}, \bibinfo {author} {\bibfnamefont {J.}~\bibnamefont {Rädler}}, \
  and\ \bibinfo {author} {\bibfnamefont {S.}~\bibnamefont {McLaughlin}},\ }\Doi
  {doi:10.1529/biophysj.106.081562} {\bibfield  {journal} {\bibinfo  {journal}
  {Biophys. J.},\ }\textbf {\bibinfo {volume} {91}},\ \bibinfo {pages} {588 }
  (\bibinfo {year} {2006})},\ ISSN \bibinfo {issn} {0006-3495}\BibitemShut
  {NoStop}%
\bibitem [{\citenamefont {Arbuzova}\ \emph {et~al.}(1998)\citenamefont
  {Arbuzova}, \citenamefont {Murray},\ and\ \citenamefont
  {McLaughlin}}]{Arbuzova1998a}%
  \BibitemOpen
  \bibfield  {author} {\bibinfo {author} {\bibfnamefont {A.}~\bibnamefont
  {Arbuzova}}, \bibinfo {author} {\bibfnamefont {D.}~\bibnamefont {Murray}}, \
  and\ \bibinfo {author} {\bibfnamefont {S.}~\bibnamefont {McLaughlin}},\ }\Doi
  {DOI: 10.1016/S0304-4157(98)00011-2} {\bibfield  {journal} {\bibinfo
  {journal} {Biochimica et Biophysica Acta (BBA) - Reviews on Biomembranes},\
  }\textbf {\bibinfo {volume} {1376}},\ \bibinfo {pages} {369 } (\bibinfo
  {year} {1998})}\BibitemShut {NoStop}%
\bibitem [{\citenamefont {Kim}\ \emph {et~al.}(1994)\citenamefont {Kim},
  \citenamefont {Shishido}, \citenamefont {Jiang}, \citenamefont {Aderem},\
  and\ \citenamefont {McLaughlin}}]{KimJ1994a}%
  \BibitemOpen
  \bibfield  {author} {\bibinfo {author} {\bibfnamefont {J.}~\bibnamefont
  {Kim}}, \bibinfo {author} {\bibfnamefont {T.}~\bibnamefont {Shishido}},
  \bibinfo {author} {\bibfnamefont {X.}~\bibnamefont {Jiang}}, \bibinfo
  {author} {\bibfnamefont {A.}~\bibnamefont {Aderem}}, \ and\ \bibinfo {author}
  {\bibfnamefont {S.}~\bibnamefont {McLaughlin}},\ }\href@noop {} {\bibfield
  {journal} {\bibinfo  {journal} {J.\ Biol.\ Chem.},\ }\textbf {\bibinfo
  {volume} {269}},\ \bibinfo {pages} {28214} (\bibinfo {year}
  {1994})}\BibitemShut {NoStop}%
\end{thebibliography}

%

\end{document}